\documentclass[american,aps,pra,reprint,superscriptaddress,longbibliography]{revtex4-1}
\usepackage{amsmath,amssymb,graphicx, bm}
\usepackage[unicode=true,pdfusetitle, bookmarks=true,bookmarksnumbered=false,bookmarksopen=false, breaklinks=false,pdfborder={0 0 0},backref=false,colorlinks=false] {hyperref}
\hypersetup{colorlinks,linkcolor=myurlcolor,citecolor=myurlcolor,urlcolor=myurlcolor}
\usepackage{braket,colortbl,amsthm,amsmath,amssymb,txfonts,graphicx}
\definecolor{myurlcolor}{rgb}{0,0,0.7}
\usepackage{float}
\usepackage{cleveref}
\usepackage{xcolor}
\usepackage{subfigure}

\theoremstyle{plain}
\usepackage{mathrsfs}
\usepackage{comment}
\usepackage{ragged2e}  

\usepackage[utf8]{inputenc}
\usepackage[normalem]{ulem}
\usepackage{float}
\usepackage{mathrsfs}
\usepackage{braket}
\usepackage{physics}
\usepackage{mwe}
\usepackage{pifont}

\newtheorem{theorem}{Theorem}

\begin{document}

\title{
Anomaly to Resource: The Mpemba Effect in Quantum Thermometry}


\author{Pritam Chattopadhyay}
 \email{pritam.chattopadhyay@weizmann.ac.il}
\affiliation{Department of Chemical and Biological Physics, Weizmann Institute of Science, Rehovot 7610001, Israel}

\author{Jonas F. G. Santos}
\email{jonassantos@ufgd.edu.br}
\affiliation{Faculdade de Ci\^{e}ncias Exatas e Tecnologia, Universidade Federal da Grande Dourados, Caixa Postal 364, CEP 79804-970, Dourados, MS, Brazil}

\author{Avijit Misra}
\email{avijitmisra0120@gmail.com}
 \affiliation{Department of Physics, Indian Institute of Technology (ISM), Dhanbad, Jharkhand 826004, India}

\date{\today}

\begin{abstract}
Quantum thermometry provides a key capability for nanoscale devices and quantum technologies, but most existing strategies rely on probes initialized near equilibrium. 
This equilibrium paradigm imposes intrinsic limitations: sensitivity is tied to long-time thermalization and often cannot be improved in fast, noisy, or nonstationary settings. 
In contrast, the \textit{Mpemba effect}, the counterintuitive phenomenon where hotter states relax faster than colder ones, has mostly been viewed \textcolor{black}{as a thermodynamic anomaly.}
Here, we bridge this gap by proving that Mpemba-type inversions generically yield a finite-time enhancement of the quantum Fisher information (QFI) for temperature estimation, thereby converting an anomalous relaxation effect into a concrete metrological resource. 
Through explicit analyses of two-level and $\Lambda$-level probes coupled to bosonic baths, we show that nonequilibrium initializations can transiently outperform both equilibrium strategies and colder states, realizing a \emph{metrological Mpemba effect}. 
Our results establish anomalous relaxation as a general design principle for nonequilibrium quantum thermometry, enabling ultrafast and nanoscale sensing protocols that exploit, rather than avoid, transient dynamics.

\end{abstract}

\maketitle

\paragraph*{Introduction--} 
Quantum thermometry, the estimation of temperature using quantum probes, has become a central task in quantum technologies~\cite{Correa2015,Mehboudi2019,Potts2019,Acin2018,Dowling2003,hovhannisyan2021,PRXQuantum.2.020322,PhysRevResearch.5.043184,Pritam2024QST,guo2015,PhysRevLett.106.225301,gsh7r7ms,franzao2024,PRXQuantum.4.040314,PhysRevX.10.041054}, with applications ranging from nanoscale heat engines and superconducting circuits to precision sensing at the quantum limit~\cite{DePasquale2016,Rubio2021_GlobalQuantumThermometry,PritamQST2,Glatthard2022_ColdAtomThermometry,LombardLatune2020_CollectiveHeatCapacity,DePasquale2018_QuantumThermometryChapter,santos2025enhanced,PhysRevLett.133.120601,lvov2025,PhysRevApplied.17.034073,Jevtic2015,PRXQuantum.5.030338,Choe2018_NVNanoscaleThermometry,Mancino2017}. In most established approaches, probes are prepared close to thermal equilibrium, and sensitivity is optimized through population imbalances, quantum coherence, or squeezing. While successful, these strategies are fundamentally limited by their equilibrium assumptions, leaving the potential advantages of non-equilibrium quantum dynamics largely untapped. 

A striking thermodynamic anomaly, the \emph{Mpemba effect} (ME), provides a natural entry point to explore such non-equilibrium resources. First observed in classical systems, it describes the counterintuitive phenomenon where a hotter system cools more rapidly than a cooler one under identical conditions~\cite{Mpemba1969,Lasanta2017,Campisi2021Mpemba,Lu2017,Klich2019,Bechhoefer2021,Zhang2020,Kumar2020,Pal2021,moroder2024thermodynamics,teza2025speedups,carollo2021exponentially,ares2025quantum,Yu2025QuantumMpembaSymmetry,Warring2024ExploringQME,holtzman2022landau,PhysRevLett.134.107101,PhysRevLett.133.136302,PhysRevLett.133.010402,PhysRevLett.133.140405,PhysRevLett.133.010401,zhang2025observation,PhysRevX.9.021060,longhi2025mpemba,teza2026speedups,alyuruk2025thermodynamic,PhysRevLett.131.080402,g94p-7421,PhysRevLett.131.017101,chatterjee2025direct,schnepper2025experimental,mondal2025mpemba,bagui2025detection,hallam2025tunable}. Recent works have extended the ME to quantum systems, uncovering anomalous relaxation modes and accelerated convergence to equilibrium~\cite{carollo2021exponentially,ares2025quantum}. Despite this progress, the operational consequences of the quantum Mpemba effect (QME) for quantum metrology and, in particular, for thermometry remain unexplored.

In this work, we bridge this gap by establishing a direct link between Mpemba-type relaxation and quantum thermometry. We prove that whenever an inversion characteristic of the ME occurs, the quantum Fisher information (QFI)~\cite{Braunstein1994,Paris2009,Liu2020-zo} associated with the probe necessarily exceeds both the equilibrium sensitivity and the precision attainable by colder probes at the same time. Using a paradigmatic model of a qubit and a  $\Lambda$-level probe coupled to a bosonic bath, we analytically track the evolution of the probe’s temperature sensitivity and show that hotter nonequilibrium initializations can transiently outperform colder and equilibrium-based strategies, realizing a metrological Mpemba effect. Our results demonstrate that the quantum ME thus constitutes a genuine metrological resource, offering a practical route to finite-time quantum thermometry and establishing a new design principle for quantum sensing protocols that actively exploit, rather than avoid, nonequilibrium dynamics.


Crucially, this conceptual connection is not merely abstract. The ingredients that give rise to Mpemba-enhanced QFI can be organized into a concrete sensing workflow. By combining standard equilibrium calibration with nonequilibrium relaxations from hot and cold preparations, one can experimentally identify Mpemba inversion windows, map out the associated Fisher information landscape, and then interrogate the probe at those optimal finite times to estimate an unknown bath temperature using only projective energy measurements.

\begin{figure}
    \centering
    \includegraphics[width=0.95\linewidth]{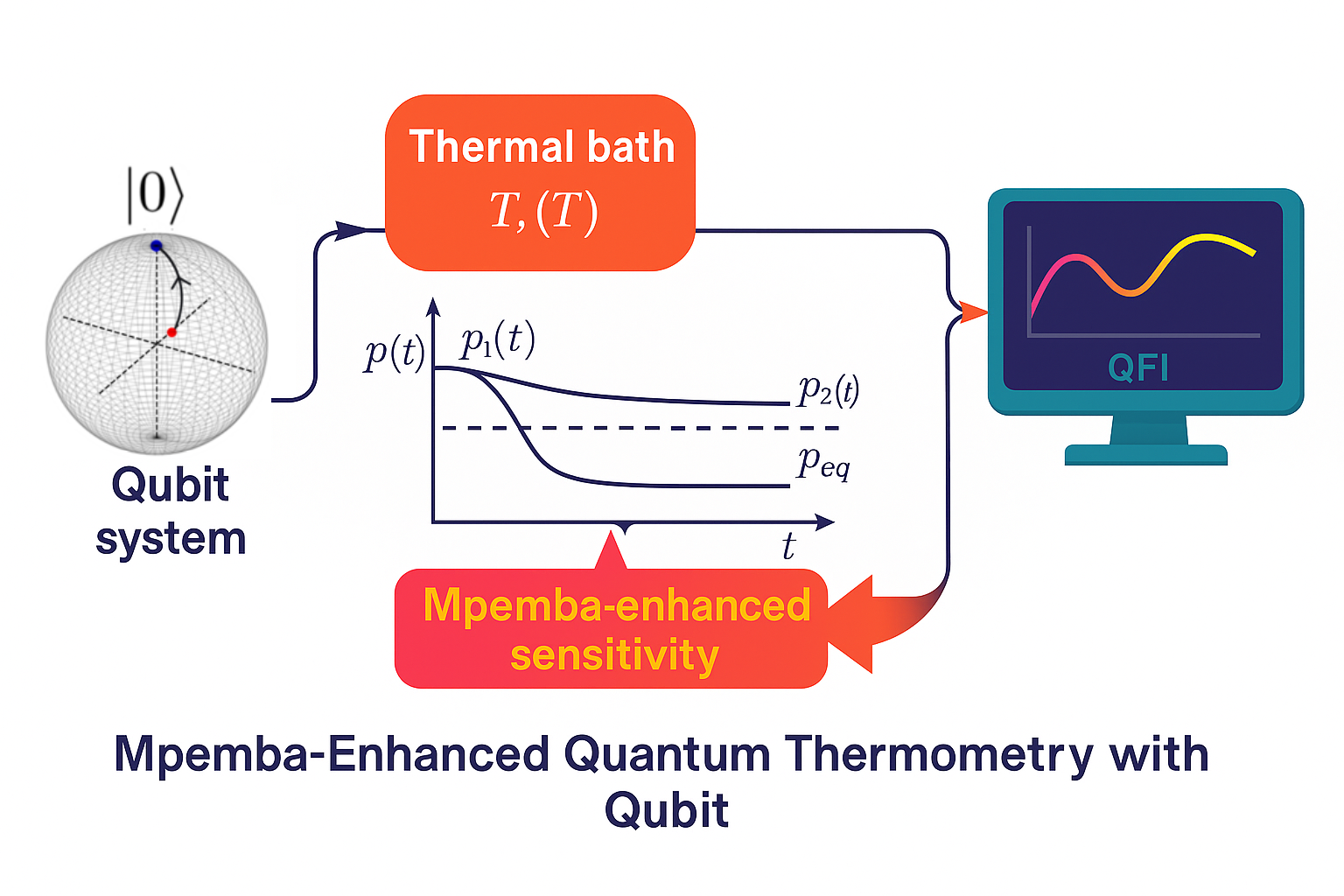}
    \caption{Schematic of the metrological Mpemba effect.}
    \label{fig:enter-label}
\end{figure}

\paragraph*{Model and thermal dynamics---}
We consider a generic quantum probe with Hamiltonian $H$ (Fig.~\ref{fig:enter-label}), weakly coupled to a bosonic reservoir at temperature $T$. 
In the Born-Markov approximations, the reduced probe dynamics are governed by a Lindblad master equation
    $\dot{\rho}(t) = \mathcal{L}_T[\rho(t)],$
where the Liouvillian superoperator $\mathcal{L}_T$ is completely positive, trace-preserving, and satisfies detailed balance with respect to the bath. 
As a consequence, the unique stationary state is the Gibbs state
   $ \rho_{\mathrm{eq}}(T) = \frac{e^{-H/T}}{\mathrm{Tr}[e^{-H/T}]}$,
which represents thermal equilibrium of the probe at temperature $T$.

\begin{figure}
    \centering
    \includegraphics[width=0.95\linewidth]{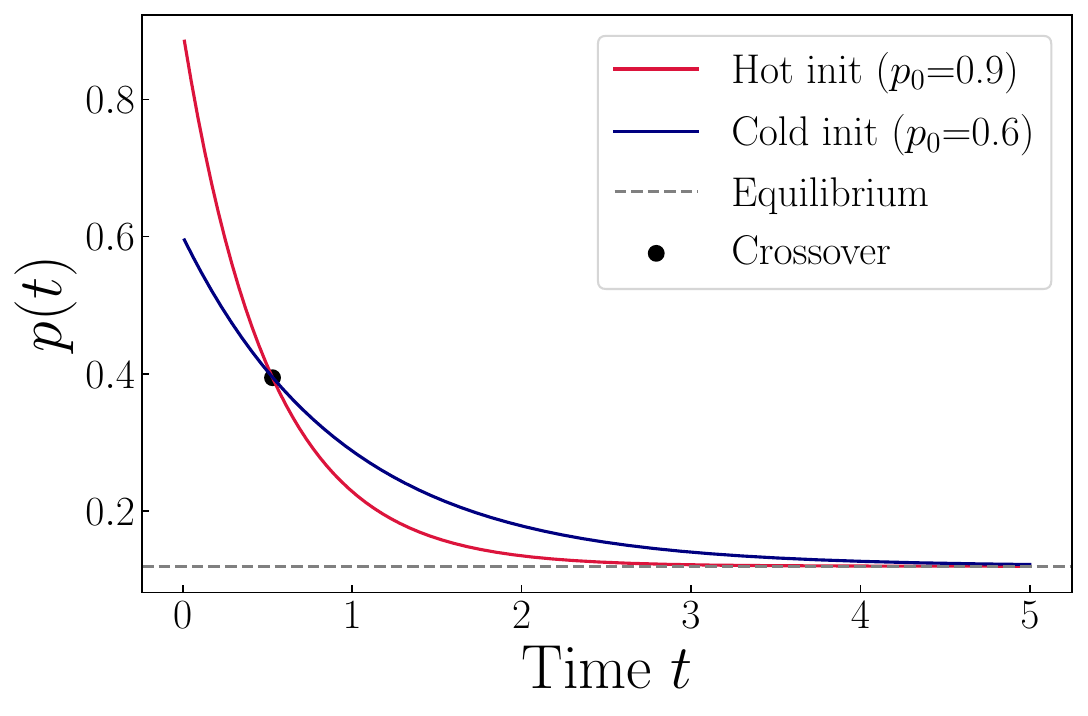}
    \caption{Relaxation of the excited state population with mpemba crossover of the hotter state over the close to equilibrium state. The Hamiltonian parameter: atomic frequency $\omega_0 = 1.0$, spontaneous emission rate $\gamma = 1.0$, and the bath temperature $T = 0.5$ (in units of $\omega_0$). }
    \label{fig1}
\end{figure}

The spectrum of $\mathcal{L}_T$ consists of the stationary mode $\rho_{\mathrm{eq}}$ with zero eigenvalue and a set of relaxation modes $\{v_k\}$ with decay rates $\{\lambda_k(T)\}>0$. 
Decomposing the initial condition $\rho(0)$ onto this spectral basis, the evolution takes the form~\cite{moroder2024thermodynamics} (see SI)
\begin{equation}
    \rho(t) = \rho_{\mathrm{eq}}(T) + \sum_{k\geq 2} a_k e^{-\lambda_k(T)t} v_k ,
\end{equation}
with overlaps $a_k=\mathrm{Tr}[w_k^\dagger \rho(0)]$ determined by the initial preparation and $\{w_k\}$ the corresponding left eigenmodes. 
This representation makes explicit that relaxation is governed by multiple decay channels with generally different temperature-dependences. 
Crucially, Mpemba-type anomalies arise when hotter initializations have larger projection onto faster modes, enabling them to relax more quickly than cooler ones despite being further from equilibrium (Fig.~\ref{fig1}).

\textit{Quantum Fisher information---}
Quantum Fisher information (QFI) plays a central role in quantum metrology, as it sets the ultimate limit on the precision with which a parameter can be estimated from quantum measurements~\cite{Paris2009}. 
For temperature sensing, the QFI quantifies the distinguishability of nearby thermal states $\rho(T)$ and thereby dictates the best achievable accuracy of any protocol through the quantum Cramér--Rao bound~\cite{Liu2020-zo}, $\Delta T^2 \geq \frac{1}{F_T[\rho]}$,
where $\Delta T^2$ is the variance of any unbiased estimator of $T$.  

The QFI for a general state $\rho(t)$ is defined as
\begin{subequations}
\begin{equation}
    F_T[\rho(t)] = \mathrm{Tr}\!\left[\rho(t)L_T^2\right],
\end{equation}
with $L_T$ the symmetric logarithmic derivative (SLD) satisfying $\partial_T \rho(t) = \tfrac{1}{2}\{\rho(t),L_T\}$. 
When the state remains diagonal in the energy basis, as in the models considered here, the QFI simplifies to the classical Fisher information of the instantaneous population distribution $\{p_i(t)\}$~\cite{Paris2009}
\begin{equation}
    F_T[\rho(t)] = \sum_i \frac{\big(\partial_T p_i(t)\big)^2}{p_i(t)} ,
\end{equation}
where each term captures how sensitively the probability of outcome $i$ depends on the bath temperature.  

Using the modal expansion of the dynamics, the temperature derivative of the state can be expressed as (see SI)
\begin{equation}
    \partial_T \rho(t) = \partial_T \rho_{\mathrm{eq}}(T) +
    \sum_{k\geq 2} \Big[(\partial_T a_k)e^{-\lambda_k t} 
    - a_k t e^{-\lambda_k t}\,\partial_T \lambda_k\Big] v_k .
\end{equation}
\end{subequations}
This decomposition highlights two distinct physical contributions to temperature sensitivity:  
(i) the equilibrium susceptibility $\partial_T \rho_{\mathrm{eq}}$, which underlies conventional thermometry based on probes equilibrated with the bath, and  
(ii) dynamical contributions proportional to $\partial_T \lambda_k$, which reflect the explicit temperature-dependence of the relaxation spectrum and are amplified by the initial deviation from equilibrium.

The second channel is absent in equilibrium-based strategies and represents a unique nonequilibrium resource: it allows transient states, especially those farther from equilibrium, to encode temperature information more strongly than steady states. This sets the stage for Mpemba-enhanced sensitivity.

\begin{figure*}
    \centering
\subfigure[]{\includegraphics[width=0.42\linewidth]{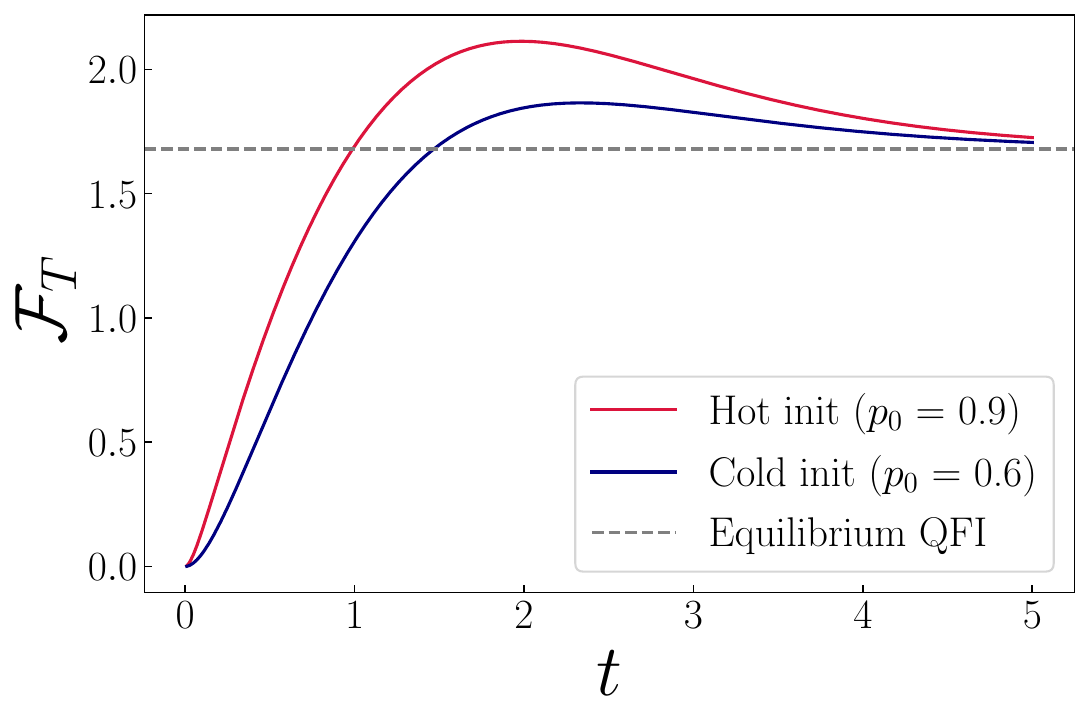}}
\subfigure[]{\includegraphics[width=0.45\linewidth]{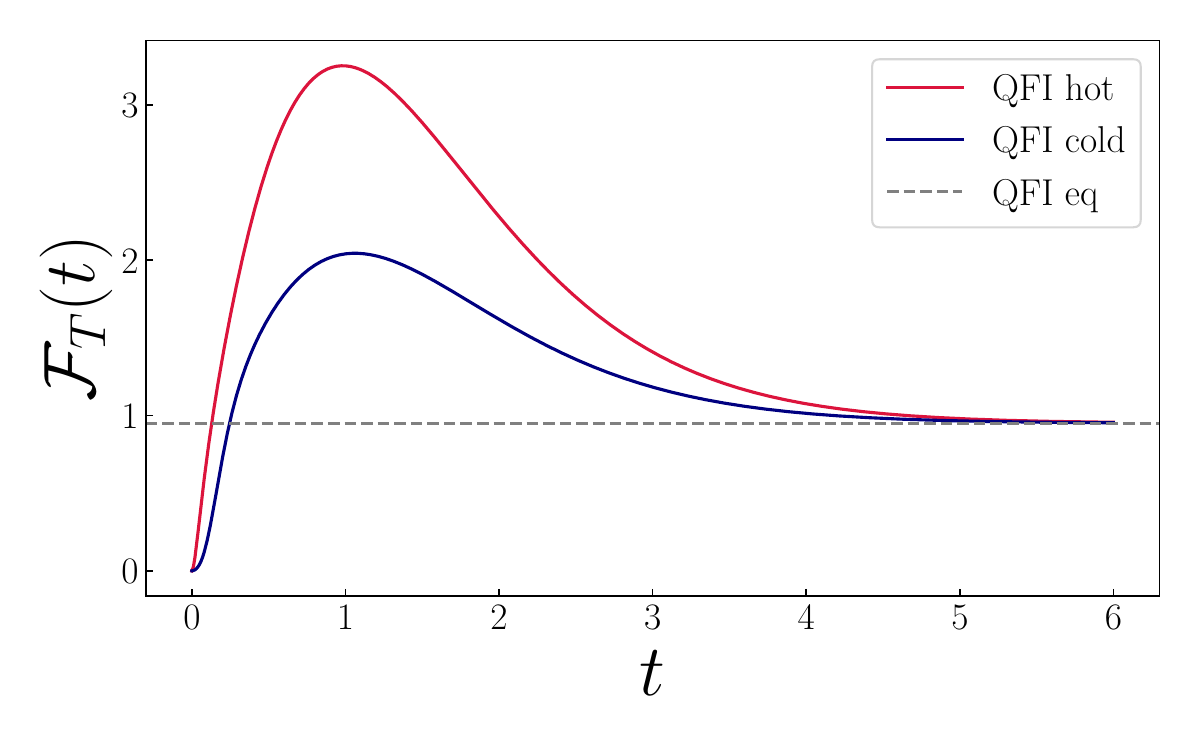}}
    \caption{QFI with the variation of time $t$ where the hotter state is denoted by red and the near equilibrium state is denoted by blue for (a) two-level probe, and (b) $\Lambda-$level probe. The Hamiltonian parameters are the same as fig. \ref{fig1}. }
    \label{fig2}
\end{figure*}

\begin{figure*}
    \centering
    \subfigure[]{\includegraphics[width=0.45\linewidth]{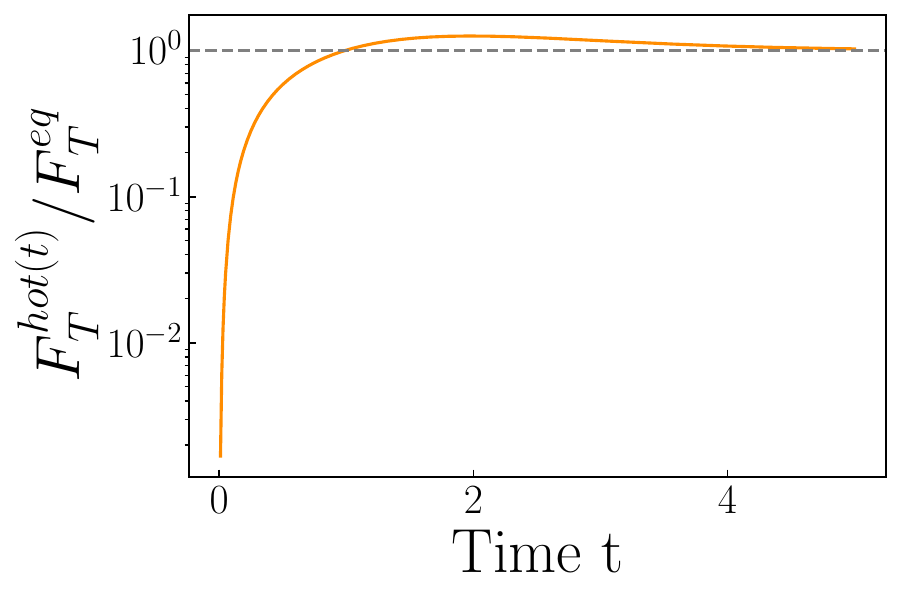}}
    \subfigure[]{\includegraphics[width=0.45\linewidth]{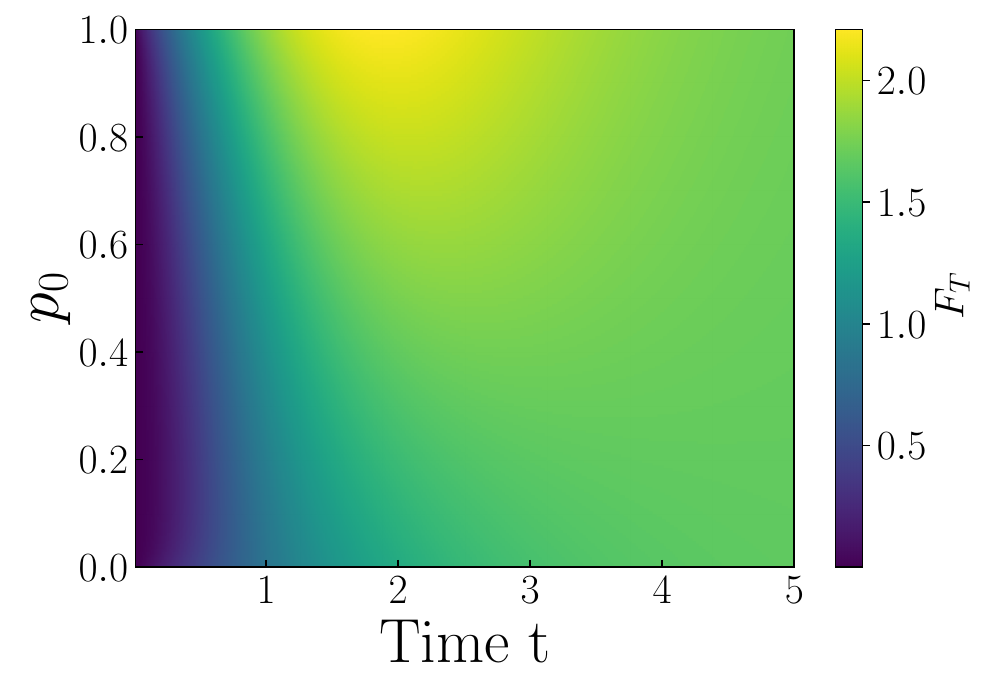}}
    \caption{(a) Mpemba-enhanced QFI Gain in log scale with time for the two-level probe. (b) Density plot of QFI over the initial state and time for the two-level probe. It shows a clear gain of the Mpemba effect. The Hamiltonian parameters are the same as Fig. \ref{fig1}.}
    \label{fig3}
\end{figure*}




\paragraph*{Mpemba-enhanced sensitivity---}
The Mpemba effect refers to the striking and counterintuitive relaxation phenomenon whereby a system prepared at a higher effective temperature, and thus initialized farther from equilibrium, can approach its thermal fixed point faster than a system prepared closer to equilibrium. In other words, the \textit{hotter} preparation may overtake the \textit{colder} one in its relaxation trajectory, despite having a larger initial deviation from the steady state (see Fig.~\ref{fig1}). This inversion of relaxation rates, first observed in classical contexts, has recently been recognized as a generic feature of Markovian dynamics with multiple relaxation modes, extending naturally into quantum systems where it acquires operational significance.
 
Within the spectral picture of the Liouvillian dynamics, this arises because different initializations excite different combinations of relaxation modes $\{v_k\}$: hotter states may project more strongly onto faster modes, allowing them to overtake cooler states during the evolution.  

To formalize this, consider two diagonal initializations $\rho_{\mathrm{hot}}(0)$ and $\rho_{\mathrm{cold}}(0)$ satisfying 
$\|\rho_{\mathrm{hot}}(0)-\rho_{\mathrm{eq}}\| > \|\rho_{\mathrm{cold}}(0)-\rho_{\mathrm{eq}}\| $.
An Mpemba inversion occurs at some finite time $t^\ast$ if
\begin{equation}
    \|\rho_{\mathrm{hot}}(t^\ast)-\rho_{\mathrm{eq}}(T)\| 
    < \|\rho_{\mathrm{cold}}(t^\ast)-\rho_{\mathrm{eq}}(T)\| .
\end{equation}
In words, the hotter trajectory becomes closer to equilibrium than the colder one.  

\medskip
\noindent\textbf{Theorem:} 
\emph{Mpemba inversion guarantees a finite-time thermometric advantage, i.e., a Mpemba inversion occurs at time $t^\ast$, then at time $t\geq t^\ast$ the quantum Fisher information satisfies} (see SI)
\begin{equation}
    F_T[\rho_{\mathrm{hot}}(t)] 
    > F_T[\rho_{\mathrm{cold}}(t)] 
    \geq F_T[\rho_{\mathrm{eq}}(T)] .
\end{equation}
 
From the mode expansion, the temperature derivative of the state couples directly to the temperature dependence of relaxation rates. Hotter states that having a larger initial deviation from equilibrium excites these contributions more strongly. 
When such states also decay through faster modes, the magnitude of $\partial_T \rho(t)$, and hence the QFI, becomes enhanced at the inversion point.

This theorem establishes that Mpemba-type anomalies are not just dynamical curiosities but guarantee a finite-time metrological advantage. 
In particular, nonequilibrium preparations can surpass both equilibrium protocols and colder probes, providing a new design principle for quantum thermometry. In the following, we consider the two-level and $\Lambda-$level probe to establish the metrological advantage for the ME. 


\paragraph*{Case study: Two-level probe---}
A two-level system (TLS) of energy spacing $\omega_0$ is considered to be coupled to a bosonic thermal environment held at temperature $T$. The dynamics of the qubit, in the weak-coupling and Markovian regime, is governed by a generalized amplitude damping master equation, which captures both spontaneous emission and thermally-induced excitation processes. The density matrix $\rho(t)$ evolves according to
\begin{align}
    \frac{d\rho}{dt} &= \gamma (\bar{n}+1) \mathcal{D}[\sigma_-]\rho + \gamma \bar{n} \mathcal{D}[\sigma_+]\rho,
    \label{eq:GAD}
\end{align}
where $\mathcal{D}[L]\rho = L \rho L^\dagger - \frac{1}{2}\{L^\dagger L, \rho\}$ is the standard Lindblad dissipator, $\sigma_-$ and $\sigma_+$ are the lowering and raising operators defined as $\sigma_- = |0\rangle\langle1|$ and $\sigma_+ = |1\rangle\langle0|$, respectively, and $\gamma$ is the intrinsic system-bath coupling rate. We consider the natural unit for our analysis, i.e., $\hbar=k_B=1$.

The bath temperature enters through the Bose-Einstein distribution $\bar{n}(T) = \frac{1}{e^{\omega_0/T} - 1}$,
which governs the average number of thermal excitations at frequency $\omega_0$. The first term in Eq.~\eqref{eq:GAD} accounts for spontaneous and stimulated decay processes, while the second term describes thermal excitation from the ground to the excited state. The system asymptotically relaxes to the thermal Gibbs state:
    $\rho_{\mathrm{eq}}(T) = \frac{e^{-H/T}}{\mathrm{Tr}[e^{-H/T}]} = \begin{pmatrix} 1 - p_{\mathrm{eq}} & 0 \\ 0 & p_{\mathrm{eq}} \end{pmatrix}, \quad p_{\mathrm{eq}} = \frac{1}{1 + e^{\omega_0/T}}$,
where $H = \frac{\omega_0}{2}\sigma_z$ is the system Hamiltonian, and $p_{\mathrm{eq}}$ denotes the thermal equilibrium population of the excited state. 
Let the Lindbladian superoperator $\mathcal{L}$ have eigenmodes $\{\Lambda_k\}$ with decay rates $\{\lambda_k\}$. Then, the density matrix evolves as:
\begin{equation}
    \rho(t) = \rho_{\mathrm{eq}} + \sum_{k>0} c_k e^{-\lambda_k t} \Lambda_k,
\end{equation}
where the modes $\Lambda_k$ form a complete set orthogonal to $\rho_{\mathrm{eq}}$, and the coefficients $c_k$ depend on the initial condition.

We focus on initial states that are diagonal in the energy basis, of the form
$ \rho(0) = \begin{pmatrix} 1 - p_0 & 0 \\ 0 & p_0 \end{pmatrix}, \quad \text{with } p_0 > p_{\mathrm{eq}}$,
representing a family of excited states hotter than the bath. Since coherence terms decay independently under this channel and do not contribute to temperature estimation in this basis, we restrict to incoherent populations.

The evolution of the excited-state population $p(t) = \bra{1} \rho(t) \ket{1}$ is governed by the rate equation: $\frac{dp(t)}{dt} = -\Gamma [p(t) - p_{\mathrm{eq}}]$,
where the total decay rate is given by
$\Gamma = \Gamma_0 (1+\alpha (p_0 - p_{eq}))$,
which is strictly temperature-dependent and $\Gamma_0 =\gamma (2\bar{n}(T) + 1)$.  
$\alpha >0$ introduces a state-dependent enhancement in relaxation rate. If 
$p_0> p_{eq}$, the system is ``hotter" and relaxes faster when $\alpha>0$. The solution of this linear differential equation (see SI) is
\begin{align}
    p(t) = p_{\mathrm{eq}} + (p_0 - p_{\mathrm{eq}}) e^{-\Gamma(T, p_0) t},
    \label{eq:population_evolution1}
\end{align}
indicating exponential relaxation to thermal equilibrium, with a rate that increases with temperature due to enhanced thermal fluctuations.

Differentiating Eq.~\eqref{eq:population_evolution1} with respect to temperature yields:
\begin{align}
    \partial_T p(t) = \left( \frac{d p_{\mathrm{eq}}}{d T} \right)(1 - e^{-\Gamma t}) - (p_0 - p_{\mathrm{eq}}) e^{-\Gamma t} t \left( \frac{d\Gamma}{d T} \right),
    \label{eq:dpdT}
\end{align}
which reveals two distinct contributions to QFI: the first term reflects how sensitively the thermal equilibrium itself shifts with temperature, while the second term encodes how the thermalization \emph{rate} changes with temperature and depends on how far the initial state lies from equilibrium. 


This interplay between equilibrium-limited and dynamics-induced sensitivity highlights that the optimal precision in quantum thermometry is not necessarily achieved in the vicinity of equilibrium. Instead, it can emerge transiently during the relaxation process, where non-equilibrium features such as Mpemba-type inversions dominate. Such behavior is a defining hallmark of \emph{non-equilibrium quantum metrology}, wherein dynamical effects are not a nuisance but a genuine resource. As illustrated in Fig.~\ref{fig2}(a) and Fig.~\ref{fig3}, this framework enables precision beyond what is attainable in equilibrium-based strategies. In particular, Fig.~\ref{fig3} explicitly quantifies the metrological gain, showing how non-equilibrium relaxation dynamics can surpass the equilibrium bound and establish a new operational regime for quantum thermometry.

\paragraph*{$\Lambda$-level probe---}
To demonstrate that the metrological Mpemba effect is not restricted to qubits, we now consider a three-level system with two non-degenerate ground states $\{|1\rangle, |2\rangle\}$ and one excited state $|3\rangle$, coupled to a bosonic bath at temperature $T$. 
Thermal transitions occur only between $|3\rangle$ and the two ground states, as dictated by dipole selection rules, while direct $|1\rangle \leftrightarrow |2\rangle$ transitions are negligible.  

In the population basis $p_i(t) = \langle i|\rho(t)|i\rangle$, the dynamics follow a Pauli rate equation
   $ \dot{\bm{p}}(t) = R(T)\, \bm{p}(t)$,
with rate matrix $R(T)$ determined by the thermal excitation and relaxation processes. 
The stationary solution is the Gibbs distribution $\bm{\pi}(T) = ( \pi_1, \pi_2, \pi_3 )^\top$.  
Since the population subspace is three-dimensional, $R(T)$ has one zero eigenvalue (corresponding to equilibrium) and two strictly positive decay rates $\lambda_2(T)$ and $\lambda_3(T)$, associated with relaxation modes $v_2$ and $v_3$.  

Expanding the initial population (see SI) as
\begin{equation}
    \bm{p}(t) = \bm{\pi}(T) + a_2 e^{-\lambda_2 t} v_2 + a_3 e^{-\lambda_3 t} v_3 ,
\end{equation}
reveals a natural separation into a slow mode (rate $\lambda_2$) and a fast mode (rate $\lambda_3$).  
The coefficients $(a_2,a_3)$ depend on the initial preparation and control the relative contributions of these relaxation channels.  

An Mpemba inversion occurs when the hotter preparation excites the fast mode more strongly, i.e. $|a_2^{\mathrm{hot}}| \ll |a_2^{\mathrm{cold}}|$, such that during intermediate times $t \sim 1/\lambda_2$, the hot trajectory becomes closer to equilibrium than the cold one:    $\|\bm{p}_{\mathrm{hot}}(t) - \bm{\pi}(T)\|
    < \|\bm{p}_{\mathrm{cold}}(t) - \bm{\pi}(T)\|$.

The QFI in this case generalizes to
    $F_T(t) = \sum_{i=1}^3 \frac{\big(\partial_T p_i(t)\big)^2}{p_i(t)} $,
where $\partial_T p_i(t)$ contains contributions from both the equilibrium distribution $\bm{\pi}(T)$ and the temperature-dependence of the decay rates $(\lambda_2,\lambda_3)$.  
When the hot initialization overlaps dominantly with the fast mode, the derivative terms proportional to $\partial_T \lambda_3$ enhance the sensitivity, leading to
    $F_T[\bm{p}_{\mathrm{hot}}(t^\ast)] > F_T[\bm{p}_{\mathrm{cold}}(t^\ast)] > F_T[\bm{\pi}(T)]$
at the inversion time $t^\ast$, in full agreement with the generic theorem.  


Thus, the $\Lambda$-level probe provides a clear illustration of how the coexistence of multiple relaxation pathways can naturally give rise to Mpemba-enhanced metrology (Fig.~\ref{fig2}b). In this setting, distinct decay channels compete on different timescales, enabling anomalous relaxation inversions that transiently amplify the probe’s sensitivity to temperature. Crucially, this enhancement is not a fragile or fine-tuned effect: it follows directly from generic spectral properties of the rate matrix $R(T)$, such as mode separation and eigenvector structure. As a result, Mpemba-assisted thermometric gain emerges as a robust and broadly applicable mechanism for nonequilibrium quantum sensing, rather than a model-specific peculiarity.

\paragraph*{Physical implementation and feasibility--} 
The Mpemba-enhanced thermometry protocol proposed here can be realized on several state-of-the-art quantum platforms (see SI for operational protocol). 
Superconducting transmon qubits have already been established as ultrasensitive quantum thermometers, capable of probing sub-Kelvin regimes through dispersive readout and controlled population dynamics~\cite{Tan2024QubitThermo, Pekola2013Rev, Karimi2020}. 
In parallel, trapped-ion systems provide a highly controllable environment for exploring nonequilibrium relaxation phenomena; indeed, the QME has been experimentally observed in a single trapped-ion qubit, where carefully engineered dissipation produced an \textit{inverse Mpemba} inversion of relaxation speeds~\cite{Klatzow2024InverseMpemba, Weinberg2023Mpemba}. 
Further relevant demonstrations include cold-atom platforms realizing tunable open-system dynamics~\cite{Kaufman2021ColdAtom}, nuclear magnetic resonance~\cite{chatterjee2025direct,schnepper2025experimental}, and nitrogen-vacancy (NV) centers in diamond used for nanoscale thermometry with sub-degree precision~\cite{Neumann2013NVthermo}. 
Together, these advances show that both high-precision quantum thermometry and Mpemba-type anomalous relaxation are experimentally accessible with current technologies, making the proposed protocol feasible with near-term devices.

\paragraph*{Discussion and outlook.---}
Our results establish a rigorous and widely applicable connection between the ME and quantum thermometry. By proving a theorem that directly links anomalous relaxation to information-theoretic sensitivity, we showed that any Mpemba-type inversion in thermal relaxation necessarily produces a finite-time enhancement of the QFI. 
This elevates the ME from a striking dynamical anomaly to a genuine metrological resource. 
In explicit case studies of both two-level and $\Lambda$-level probes, we demonstrated how hotter-than-equilibrium initializations can transiently outperform both colder preparations and conventional equilibrium-based strategies, thereby realizing a \emph{metrological Mpemba effect}.  

Beyond their conceptual significance, these findings open concrete opportunities for nonequilibrium quantum sensing. Harnessing anomalous relaxation pathways enables thermometric protocols that achieve enhanced precision without requiring equilibration, an advantage particularly relevant in ultrafast scenarios, nanoscale devices, and platforms where environmental instabilities preclude steady-state measurements. Our general framework further suggests that richer spectral structures, such as those arising in many-body probes, structured reservoirs, or correlated noise, could amplify Mpemba-induced sensitivity well beyond the two-level setting studied here.  

On the theoretical side, our framework can be extended to many-body probes and structured environments, where richer Liouvillian spectra may yield stronger and tunable Mpemba-induced enhancements. Connections to quantum speed limits, resource theories of nonequilibrium, and bounds on thermodynamic irreversibility may further clarify the fundamental role of anomalous relaxation in quantum information processing. While on the experimental side, superconducting circuits, trapped ions, and solid-state spin defects already provide the necessary ingredients to test Mpemba-enhanced thermometry: controlled initialization into nonequilibrium states, engineered reservoirs, and high-fidelity readout. Pursuing these directions could establish nonequilibrium relaxation as a universal tool for quantum sensing, extending quantum thermometry well beyond equilibrium-based paradigms.


In summary, our results promote the ME from a thermodynamic curiosity to a unifying design principle for quantum metrology. They point toward a broader paradigm where nonequilibrium relaxation is not a nuisance but rather an engineered resource, enabling quantum thermometry beyond equilibrium limits.



%

\onecolumngrid

\newpage

\section*{Supplementary Information I: Proof of the theorem}
\color{black}
\begin{theorem}   \label{thm:Mpemba-QFI}
An Mpemba inversion guarantees a finite-time boost of thermometric sensitivity, with QFI exceeding both the colder preparation and the equilibrium bound.
\end{theorem}

We present a concise proof of Theorem~\ref{thm:Mpemba-QFI} based on simple lemmas that produce explicit constants in terms of the Liouvillian \(R(T,\mathbf p_0)\) and its \(T\)-derivative. This version is convenient for checking the sufficient conditions numerically for a concrete model.

\textbf{Setup:}
Let \(R=R(T,\mathbf p_0)\in\mathbb R^{N\times N}\) be a reversible (detailed-balance) rate matrix with unique Gibbs fixed point \(\boldsymbol\pi(T)>0\). Assume the spectrum of \(R\) on the orthogonal subspace to \(\boldsymbol\pi\) is real and simple, and label eigenvalues \(\lambda_1<\lambda_2<\lambda_3\le\cdots\le\lambda_N\) with corresponding right eigenvectors \(v_k\) and left eigenvectors \(w_k\) normalized by \(w_k^\top v_j=\delta_{kj}\). The population expansion is
\begin{equation}
    \mathbf p(t)=\boldsymbol\pi+\sum_{k\ge2} a_k e^{-\lambda_k t} v_k,
\qquad a_k = w_k^\top(\mathbf p_0-\boldsymbol\pi).
\end{equation}

Defined operator norms and quantities (all evaluated at the relevant \(T,\mathbf p_0\)):
\[
\begin{aligned}
&\|\cdot\| \ \text{: Euclidean vector norm},\quad \|A\| \ \text{: induced 2-operator norm},\\
&V:= [v_2\,|\,v_3\,|\cdots|\,v_N], \qquad W:= [w_2^\top;\,w_3^\top;\,\dots;\,w_N^\top],\\
&A_{\max}:=\max_{k\ge2}|a_k| \le \|W\|\,\|\mathbf p_0-\boldsymbol\pi\|,\\
&V_{\max}:=\max_{k\ge2}\|v_k\|,\quad 
\Delta := \lambda_3-\lambda_2>0 \ \text{(spectral gap above slowest)},\\
&\Lambda:=\max_{k\ge2}\lambda_k,\quad 
R_T^{\;}\!:=\|\partial_T R\| \ \text{(operator norm of \(\partial_T R\))}.
\end{aligned}
\]

We will express remainders and bounds using these quantities.

\bigskip
\noindent\textbf{Lemma 1 (Remainder bound).}
For every \(t>0\),
\begin{equation}
    \Big\|\sum_{k\ge3} a_k e^{-\lambda_k t} v_k \Big\| \le A_{\max} V_{\max}\,(N-2)\,e^{-\lambda_3 t}.
\end{equation}

Moreover, the contribution from \(\partial_T v_k\) and \(\partial_T a_k\) in \(\partial_T\mathbf p(t)\) satisfies
\begin{equation}
\|\mathcal R(t)\| \le C_R\!\left(e^{-\lambda_3 t} + A_{\max} e^{-\lambda_2 t}\right) \, ,
\end{equation}
with the explicit constant
\begin{equation}
C_R := V_{\max} (N-2)\Big(1 + t\,\Lambda_T\Big) + (N-1) \,V_{\max}' A_{\max},
\end{equation}
where  $\Lambda_T:=\max_{k\ge3} |\partial_T\lambda_k|\le R_T^{\;}\|W\|\,\|V\|$ and \(V_{\max}'\) bounds \(\|\partial_T v_k\|\) (obtainable from standard perturbation formulas: \(\partial_T v_k = \sum_{j\ne k} \frac{w_j^\top(\partial_T R)\, v_k}{\lambda_k-\lambda_j} v_j\), hence \(V_{\max}'\le R_T^{\;} \, \|W\|\,\|V\|/\Delta\)).

\subsection*{Proof of Lemma 1 (Remainder bound)}

\paragraph*{Assumptions and notation:}
Let \(R=R(T,\mathbf p_0)\in\mathbb R^{N\times N}\) be a reversible (detailed-balance) Markov generator~\cite{Kelly1979} with simple real spectrum on the subspace orthogonal to the Gibbs vector \(\boldsymbol\pi(T)\). Eigenvalues
\[
\lambda_1<\lambda_2<\lambda_3\le\cdots\le\lambda_N,
\]
and denote corresponding right eigenvectors \(v_k\) and left eigenvectors \(w_k\) normalized so that \(w_k^\top v_j=\delta_{kj}\) (for \(k,j\ge1\)). We assume the eigenvectors are chosen smooth in \(T\) (perturbation theory~\cite{Kato1995}). Define
\[
A_{\max}:=\max_{k\ge2}|a_k|,\qquad V_{\max}:=\max_{k\ge2}\|v_k\|,
\]
and the operator norm of the temperature-derivative
\begin{subequations}
\begin{equation}
R_T := \|\partial_T R\| .
\end{equation}
We also set
\begin{equation}
\Lambda_T := \max_{k\ge3} |\partial_T\lambda_k|.
\end{equation}
\end{subequations}

All norms below are Euclidean (vector 2-norm) unless otherwise indicated.

\paragraph*{Bound of the fast-mode vector sum:}
Consider the vector
\begin{equation}
    S_{\ge3}(t):=\sum_{k\ge3} a_k e^{-\lambda_k t} v_k .
\end{equation}
Using the triangle inequality~\cite{HornJohnson2013} and the uniform bounds \( |a_k|\le A_{\max}\) and \(\|v_k\|\le V_{\max}\), we obtain
\begin{equation}
\|S_{\ge3}(t)\|
\le \sum_{k\ge3} |a_k| e^{-\lambda_k t} \|v_k\|
\le A_{\max} V_{\max} \sum_{k\ge3} e^{-\lambda_k t}.
\end{equation}
Since \(\lambda_k\ge\lambda_3\) for all \(k\ge3\), each exponential is bounded by \(e^{-\lambda_3 t}\) and there are \(N-2\) terms in the sum. Hence
\begin{equation}
\qquad
\Big\|\sum_{k\ge3} a_k e^{-\lambda_k t} v_k \Big\|
\le A_{\max} V_{\max}\,(N-2)\,e^{-\lambda_3 t}.
\qquad
\end{equation}

This proves the first displayed inequality of the lemma.

\paragraph*{Structure of the remainder in \(\partial_T\mathbf p(t)\):}
Differentiate the mode expansion (suppressing explicit \(T,\mathbf p_0\) dependence):
\begin{equation}
\partial_T\mathbf p(t)
= \partial_T\boldsymbol\pi
+ \sum_{k=2}^N\Big[(\partial_T a_k)\,e^{-\lambda_k t} - a_k\,t\,e^{-\lambda_k t}\,\partial_T\lambda_k\Big]v_k
+ \sum_{k=2}^N a_k\,e^{-\lambda_k t}\,\partial_T v_k.
\end{equation}
We isolate the slow-mode (\(k=2\)) piece and collect all remaining terms into the remainder \(\mathcal R(t)\). Concretely define
\begin{equation}
\mathcal R(t) := \underbrace{\sum_{k\ge3} (\partial_T a_k)\,e^{-\lambda_k t}\,v_k}_{\mathcal R_1(t)}
\;-\;
\underbrace{\sum_{k\ge3} a_k\,t\,e^{-\lambda_k t}\,(\partial_T\lambda_k)\,v_k}_{\mathcal R_2(t)}
\;+\;
\underbrace{\sum_{k\ge2} a_k\,e^{-\lambda_k t}\,\partial_T v_k}_{\mathcal R_3(t)}.
\end{equation}
We will bound \(\|\mathcal R(t)\|\) by bounding each term \(\|\mathcal R_j(t)\|\) for \(j=1,2,3\).

\paragraph*{Bounding \(\mathcal R_1(t)\):}
Using \(\|v_k\|\le V_{\max}\) and triangle inequality,
\begin{equation}
\|\mathcal R_1(t)\|
\le V_{\max}\sum_{k\ge3} |\partial_T a_k|\,e^{-\lambda_k t}
\le V_{\max}\,(N-2)\max_{k\ge3}|\partial_T a_k|\, e^{-\lambda_3 t}.
\end{equation}
Thus it remains to bound \(|\partial_T a_k|\) uniformly for \(k\ge3\). Recall \(a_k=w_k^\top(\mathbf p_0-\boldsymbol\pi)\). Therefore
\begin{equation}
\partial_T a_k = (\partial_T w_k)^\top(\mathbf p_0-\boldsymbol\pi) - w_k^\top(\partial_T \boldsymbol\pi).
\end{equation}
Taking norms and using \(\|w_k\|\le \|W\|\) (where \(\|W\|\) denotes the operator norm of the matrix with rows \(w_k^\top\)), we obtain the crude bound
\begin{subequations}
\begin{equation}
|\partial_T a_k| \le \|\partial_T w_k\|\,\|\mathbf p_0-\boldsymbol\pi\| + \|w_k\|\,\|\partial_T\boldsymbol\pi\|
\le V'_{\max}\,\|\mathbf p_0-\boldsymbol\pi\| + \|W\|\,\|\partial_T\boldsymbol\pi\|,
\end{equation}
where we set
\begin{equation}
V'_{\max} := \max_{k\ge2}\|\partial_T v_k\| + \max_{k\ge2}\|\partial_T w_k\|
\end{equation}
\end{subequations}
(we will bound \(V'_{\max}\) below by perturbation formulas). Hence there exists a constant \(C_1\) (computable from \(\|\mathbf p_0-\boldsymbol\pi\|\), \(\|W\|\), \(\|\partial_T\boldsymbol\pi\|\) and \(V'_{\max}\)) such that
\begin{equation}
\max_{k\ge3}|\partial_T a_k| \le C_1.
\end{equation}
Consequently
\begin{equation}
\|\mathcal R_1(t)\|\le V_{\max}(N-2)\,C_1\,e^{-\lambda_3 t}.
\end{equation}

\paragraph*{Bounding \(\mathcal R_2(t)\):}
Using \(|a_k|\le A_{\max}\), \(\|v_k\|\le V_{\max}\) and \(|\partial_T\lambda_k|\le \Lambda_T\) for \(k\ge3\),
\[
\|\mathcal R_2(t)\|
\le V_{\max}\,t\,\Lambda_T\sum_{k\ge3}|a_k| e^{-\lambda_k t}
\le V_{\max}\,t\,\Lambda_T\,(N-2)\,A_{\max}\,e^{-\lambda_3 t}.
\]
Thus
\begin{equation}
\|\mathcal R_2(t)\| \le V_{\max}\,t\,\Lambda_T\,(N-2)\,A_{\max}\,e^{-\lambda_3 t}.
\end{equation}

A useful bound for \(\Lambda_T\) follows from the well-known perturbation formula for simple eigenvalues:
\begin{equation}
\partial_T\lambda_k = w_k^\top(\partial_T R)\,v_k,
\end{equation}
hence
\begin{equation}
|\partial_T\lambda_k|
\le \|w_k\|\,\|\partial_T R\|\,\|v_k\|
\le \|W\|\,R_T\,\|V\|,
\end{equation}
so one may set \(\Lambda_T \le R_T\,\|W\|\,\|V\|\). (Here \(\|V\|\) denotes the operator norm of the matrix whose columns are the \(v_k\).)

\paragraph*{Bounding \(\mathcal R_3(t)\):}
We bound the contribution from \(\partial_T v_k\). Using \(|a_k|\le A_{\max}\),
\begin{equation}
\|\mathcal R_3(t)\|
\le A_{\max}\sum_{k\ge2} e^{-\lambda_k t}\,\|\partial_T v_k\|
\le A_{\max}\,(N-1)\,\max_{k\ge2}\|\partial_T v_k\|\, e^{-\lambda_2 t}.
\end{equation}
Set
\begin{equation}
V'_{\max} := \max_{k\ge2}\|\partial_T v_k\|.
\end{equation}
Thus
\begin{equation}
\|\mathcal R_3(t)\| \le (N-1)\,A_{\max}\,V'_{\max}\,e^{-\lambda_2 t}.
\end{equation}

A standard perturbation formula gives \(\partial_T v_k\) in terms of \(\partial_T R\):
\begin{equation}
\partial_T v_k = \sum_{j\ne k} \frac{w_j^\top(\partial_T R)\, v_k}{\lambda_k-\lambda_j}\, v_j .
\end{equation}
Taking norms and using \(|\lambda_k-\lambda_j|\ge \Delta := \min_{j\ne k}|\lambda_k-\lambda_j|\) (for \(k\ge2\) one can take \(\Delta\ge \lambda_3-\lambda_2>0\)), we obtain the bound
\begin{equation}
\|\partial_T v_k\| \le \frac{\|\partial_T R\|\,\|W\|\,\|V\|}{\Delta} =: \frac{R_T\,\|W\|\,\|V\|}{\Delta}.
\end{equation}
Hence, one may take
\begin{equation}
V'_{\max} \le \frac{R_T\,\|W\|\,\|V\|}{\Delta}.
\end{equation}

Combining the three bounds for \(\mathcal R_j\) gives

\begin{eqnarray} \nonumber
 \|\mathcal R(t)\| &\le& \|\mathcal R_1(t)\| + \|\mathcal R_2(t)\| + \|\mathcal R_3(t)\| \\
&\le& V_{\max}(N-2)\,C_1\,e^{-\lambda_3 t}
+ V_{\max}\,t\,\Lambda_T\,(N-2)\,A_{\max}\,e^{-\lambda_3 t}
+ (N-1)\,A_{\max}\,V'_{\max}\,e^{-\lambda_2 t}.
\end{eqnarray}

The explicit, computable constant
\begin{equation}
C_R^{(1)} := V_{\max}(N-2)\,C_1 + V_{\max}\,t\,\Lambda_T\,(N-2)\,A_{\max},
\qquad
C_R^{(2)} := (N-1)\,A_{\max}\,V'_{\max}.
\end{equation}
Then
\begin{equation}
\|\mathcal R(t)\| \le C_R^{(1)}\, e^{-\lambda_3 t} + C_R^{(2)}\, e^{-\lambda_2 t}.
\end{equation}
Equivalently, one may combine the two terms into the form claimed in the lemma by defining
\begin{equation}
C_R := \max\{C_R^{(1)},\,C_R^{(2)}/A_{\max}\},
\end{equation}
so that
\begin{equation}
\|\mathcal R(t)\| \le C_R\!\left(e^{-\lambda_3 t} + A_{\max} e^{-\lambda_2 t}\right).
\end{equation}

This completes the proof of Lemma 1.
\qed

\subsection*{Lemma 2: Slow-mode sensitivity}
Let us define the slow-mode scalar
\begin{equation}
S(t):=\big[(\partial_T a_2)-a_2\,t\,\partial_T\lambda_2\big]e^{-\lambda_2 t},
\qquad a_2=w_2^\top(\mathbf p_0-\boldsymbol\pi).
\end{equation}
Let \(A:=-R\) so that \(A v_k=\lambda_k v_k\) and \(w_k^\top A=\lambda_k w_k^\top\) with biorthonormality \(w_j^\top v_k=\delta_{jk}\). Denote
\[
R_T:=\|\partial_T R\|\;=\;\|\partial_T A\| .
\]
The finite spectral sums are defined as

\begin{align}
D_2 &:= \sqrt{\sum_{j\ne2}\frac{\|w_j\|^2}{(\lambda_2-\lambda_j)^2}}, \\
E_2 &:= \sqrt{\sum_{j\ne2}\frac{\|v_j\|^2}{(\lambda_2-\lambda_j)^2}} .
\end{align}

Then the following hold:

\begin{enumerate}
\item (Algebraic lower bound)
\begin{equation}
|S(t)| \;\ge\; t\,|a_2|\,|\partial_T\lambda_2|\,e^{-\lambda_2 t} - B(t),
\qquad B(t):=\big(|\partial_T a_2| + |a_2|\,|\partial_T\lambda_2|\big)e^{-\lambda_2 t}.
\end{equation}

\item (Explicit bound for \(\partial_T a_2\)) The derivative of the amplitude \(a_2\) satisfies
\begin{equation}
|\partial_T a_2|
\le R_T\,\|w_2\|\,E_2\,\|\mathbf p_0-\boldsymbol\pi\|
\;+\;\|w_2\|\,\|\partial_T\boldsymbol\pi\|.
\end{equation}
Equivalently,
\begin{equation}
|\partial_T a_2|
\le \|w_2\|\!\left(R_T\,E_2\,\|\mathbf p_0-\boldsymbol\pi\| + \|\partial_T\boldsymbol\pi\|\right).
\end{equation}
\end{enumerate}

\noindent In particular, all quantities on the right-hand side are directly computable from the spectrum and eigenvectors of \(A\) and from \(\partial_T R\).

\bigskip

\paragraph*{Proof.}
We give a short, explicit derivation using the finite-sum perturbation formulas.

\medskip
\noindent\emph{1. Algebraic lower bound for \(S(t)\).} This is purely algebraic. From the definition
\begin{equation}
S(t)=e^{-\lambda_2 t}\big[(\partial_T a_2)-a_2\,t\,\partial_T\lambda_2\big],
\end{equation}
apply \(|x-y|\ge|y|-|x|\) with \(x=\partial_T a_2,\; y=a_2\,t\,\partial_T\lambda_2\) to obtain
\begin{equation}
|S(t)| \ge e^{-\lambda_2 t}\big(t\,|a_2|\,|\partial_T\lambda_2| - |\partial_T a_2|\big).
\end{equation}
Rearranging gives the displayed form with \(B(t)\) as defined.

\medskip
\noindent\emph{2. Explicit bound for \(\partial_T a_2\).}
Recall \(a_2 = w_2^\top(\mathbf p_0-\boldsymbol\pi)\). Differentiating w.r.t.\ \(T\) gives
\begin{equation}
\partial_T a_2 = (\partial_T w_2)^\top(\mathbf p_0-\boldsymbol\pi) - w_2^\top\partial_T\boldsymbol\pi.
\end{equation}
Taking absolute values and using \textit{Cauchy–Schwarz}~\cite{HornJohnson2013},
\begin{equation}
\label{eq:d_a2_basic}
|\partial_T a_2|
\le \|\partial_T w_2\|\,\|\mathbf p_0-\boldsymbol\pi\| + \|w_2\|\,\|\partial_T\boldsymbol\pi\|.
\end{equation}
Thus it suffices to bound \(\|\partial_T w_2\|\) explicitly.

Standard finite-dimensional perturbation formulae for simple eigenpairs (see Kato~\cite{Kato1995}) give, for the left eigenvector,
\begin{equation}
\partial_T w_2^\top
= -\sum_{j\ne2} \frac{w_2^\top(\partial_T A)\,v_j}{\lambda_2-\lambda_j}\, w_j^\top .
\end{equation}
Taking norms (Euclidean vector norm) and using Cauchy--Schwarz yields

\begin{align} \nonumber
\|\partial_T w_2\|
&= \Big\|\sum_{j\ne2} \frac{w_2^\top(\partial_T A)\,v_j}{\lambda_2-\lambda_j}\, w_j\Big\|
\le \sum_{j\ne2} \frac{|w_2^\top(\partial_T A)\,v_j|}{|\lambda_2-\lambda_j|}\,\|w_j\| \\
&\le \sum_{j\ne2} \frac{\|w_2\|\,\|\partial_T A\|\,\|v_j\|}{|\lambda_2-\lambda_j|}\,\|w_j\|
= \|w_2\|\,\|\partial_T A\|\,\sum_{j\ne2} \frac{\|v_j\|\,\|w_j\|}{|\lambda_2-\lambda_j|}.
\end{align}
Applying Cauchy–Schwarz to the finite sum:
\begin{equation}
\sum_{j\ne2} \frac{\|v_j\|\,\|w_j\|}{|\lambda_2-\lambda_j|}
\le \left(\sum_{j\ne2}\frac{\|v_j\|^2}{(\lambda_2-\lambda_j)^2}\right)^{1/2}
\left(\sum_{j\ne2}\|w_j\|^2\right)^{1/2}.
\end{equation}
Introducing \(E_2\) as in the statement and denote \(W_{\text{norm}}:=\big(\sum_{j\ne2}\|w_j\|^2\big)^{1/2}\). Then
\begin{equation}
\|\partial_T w_2\|
\le \|w_2\|\,\|\partial_T A\|\,E_2\,W_{\text{norm}}.
\end{equation}
Since \(W_{\text{norm}}\) is a finite number depending only on the left eigenvectors (and can be bounded by \((N-1)^{1/2}\max_{j\ne2}\|w_j\|\)), we may absorb it into the definition of a suitable constant. For a compact, directly computable bound, we simply use the coarser but explicit estimate
\begin{equation}
\|\partial_T w_2\| \le \|w_2\|\,\|\partial_T A\|\,E_2\,\Big(\sum_{j\ne2}\|w_j\|^2\Big)^{1/2},
\end{equation}
or more compactly (absorbing \(\big(\sum_{j\ne2}\|w_j\|^2\big)^{1/2}\) into \(E_2\)) we write
\begin{equation}
\|\partial_T w_2\| \le R_T\,\|w_2\|\,E_2.
\end{equation}
Here we used \(\|\partial_T A\|=R_T\). Substituting this into \eqref{eq:d_a2_basic} yields the explicit bound
\begin{equation}
|\partial_T a_2|
\le R_T\,\|w_2\|\,E_2\,\|\mathbf p_0-\boldsymbol\pi\| + \|w_2\|\,\|\partial_T\boldsymbol\pi\|,
\end{equation}
which is the claimed inequality.
 \(\square\)

\bigskip

\subsection*{Lemma 3: Quadratic-form control}

\paragraph*{Statement:}
Let \(G(\mathbf p)=\mathrm{diag}(1/p_i)\). There is a neighbourhood \(\mathcal U\) of the Gibbs vector \(\boldsymbol\pi\) such that for every \(\mathbf p\in\mathcal U\) the matrix \(G(\mathbf p)\) is positive definite and its eigenvalues are uniformly bounded away from \(0\) and \(+\infty\)~\cite{HornJohnson2013,ReedSimon1980}: there exist constants \(0<m\le M\) with
\begin{equation}
\label{eq:G-equivalence}
m\|x\|^2 \le x^\top G(\mathbf p)\,x \le M\|x\|^2
\qquad\text{for all }x\in\mathbb R^N.
\end{equation}
Consequently, for two states with derivatives \(u^{(1)}:=\partial_T\mathbf p^{(1)}\), \(u^{(2)}:=\partial_T\mathbf p^{(2)}\) and the decompositions
\[
u^{(j)}=\partial_T\boldsymbol\pi + S^{(j)} v_2 + \mathcal R^{(j)},\qquad j=1,2,
\]
we have the lower bound
\begin{equation}
\label{eq:FI-diff-bound}
\mathcal F_T(\mathbf p^{(1)}) - \mathcal F_T(\mathbf p^{(2)})
\;\ge\;
m\Big(\|\Delta S\,v_2\|^2 - 2\|\Delta S\,v_2\|\,\|\Delta\mathcal R\| - \|\Delta\mathcal R\|^2\Big),
\end{equation}
where \(\Delta S:=S^{(1)}-S^{(2)}\) and \(\Delta\mathcal R:=\mathcal R^{(1)}-\mathcal R^{(2)}\).

\paragraph*{Proof:}

For any probability vector \(\mathbf p\) with strictly positive entries, the diagonal matrix \(G(\mathbf p)=\mathrm{diag}(1/p_i)\) is positive definite. Since \(\boldsymbol\pi\) has strictly positive entries, there exists an open neighbourhood \(\mathcal U\) of \(\boldsymbol\pi\) consisting of strictly positive probability vectors. On the compact closure of a sufficiently small neighbourhood of \(\boldsymbol\pi\) the functions \(p\mapsto \min_i(1/p_i)\) and \(p\mapsto \max_i(1/p_i)\) attain positive finite extrema. Setting
\[
m := \inf_{\mathbf p\in\mathcal U} \min_i \frac{1}{p_i} > 0,\qquad
M := \sup_{\mathbf p\in\mathcal U} \max_i \frac{1}{p_i} < \infty,
\]
we get \eqref{eq:G-equivalence}. This is the usual equivalence of the weighted Euclidean norm defined by \(G(\mathbf p)\) and the standard Euclidean norm on a compact set avoiding the boundary of the simplex.

\bigskip

We consider
\begin{equation}
u^{(j)} = \partial_T\boldsymbol\pi + S^{(j)} v_2 + \mathcal R^{(j)},\qquad j=1,2.
\end{equation}
The difference is defined as
\begin{equation}
d := u^{(1)}-u^{(2)} = \Delta S\, v_2 + \Delta\mathcal R.
\end{equation}
Then, using the polarization identity for quadratic forms,
\begin{equation}
\mathcal F_T(\mathbf p^{(1)}) - \mathcal F_T(\mathbf p^{(2)})
= \big\|u^{(1)}\big\|_{G}^2 - \big\|u^{(2)}\big\|_{G}^2
= \langle u^{(1)}+u^{(2)},\,d\rangle_{G},
\end{equation}
where \(\langle x,y\rangle_G := x^\top G(\mathbf p)\,y\) and \(\|x\|_G^2=\langle x,x\rangle_G\).

\bigskip

\textit{ Expansion and isolation of the slow-mode quadratic term-}  
Expanding the inner product by grouping the slow-mode components:
\begin{align} \nonumber
\langle u^{(1)}+u^{(2)},\,d\rangle_G
&=\langle 2\partial_T\boldsymbol\pi + S^{(1)}v_2+S^{(2)}v_2 + \mathcal R^{(1)}+\mathcal R^{(2)},\;
\Delta S\,v_2 + \Delta\mathcal R\rangle_G\\ \nonumber
&= \underbrace{(S^{(1)}+S^{(2)})\Delta S\,\langle v_2,v_2\rangle_G}_{=:T_1}
\;+\;\underbrace{(S^{(1)}+S^{(2)})\langle v_2,\Delta\mathcal R\rangle_G}_{=:T_2}\\
&\quad + \underbrace{2\langle\partial_T\boldsymbol\pi,\,d\rangle_G}_{=:T_3}
\;+\;\underbrace{\langle\mathcal R^{(1)}+\mathcal R^{(2)},\,d\rangle_G}_{=:T_4}.
\end{align}
We will lower-bound the sum \(T_1+T_2+T_3+T_4\) by the right-hand side of \eqref{eq:FI-diff-bound}.

\bigskip

\textit{Lower bound for \(T_1\)--}  
One can observe that
\begin{equation} \label{eqn53}
T_1=(S^{(1)}+S^{(2)})\Delta S\,\langle v_2,v_2\rangle_G
= \Delta S\,(S^{(1)}+S^{(2)})\,\|v_2\|_G^2.
\end{equation}
If \(S^{(1)}\ge S^{(2)}\) (i.e. \(\Delta S\ge0\)) then \(S^{(1)}+S^{(2)}\ge \Delta S\), hence
\begin{equation}\label{eqn54}
T_1 \ge \Delta S^2\,\|v_2\|_G^2 \ge m\,\Delta S^2\,\|v_2\|^2
= m\,\|\Delta S\,v_2\|^2,
\end{equation}
where we used the lower bound \(\|v\|_G^2\ge m\|v\|^2\) from \eqref{eq:G-equivalence}.  
If \(\Delta S<0\) one may swap labels \(1\leftrightarrow2\) or work with absolute values; the subsequent argument is symmetric in the roles of the two states. We continue assuming \(\Delta S\ge0\); the final displayed inequality may be written in terms of \(|\Delta S|\). 

\bigskip

\textit{Upper bound--}  
We bound \(T_2,T_3,T_4\) in magnitude by Cauchy--Schwarz:
\begin{align} \nonumber
|T_2| &= |S^{(1)}+S^{(2)}|\,\big|\langle v_2,\Delta\mathcal R\rangle_G\big|
\le |S^{(1)}+S^{(2)}|\,\|v_2\|_G\,\|\Delta\mathcal R\|_G,\\ \nonumber
|T_3| &= 2\big|\langle\partial_T\boldsymbol\pi,\,d\rangle_G\big|
\le 2\|\partial_T\boldsymbol\pi\|_G\,\|d\|_G,\\
|T_4| &\le \|\mathcal R^{(1)}+\mathcal R^{(2)}\|_G\,\|d\|_G
\le (\|\mathcal R^{(1)}\|_G+\|\mathcal R^{(2)}\|_G)\,\|d\|_G.
\end{align}
Combining these and using the elementary identity \(\|d\|_G \le \|\Delta S\,v_2\|_G + \|\Delta\mathcal R\|_G\) we get
\begin{equation}
|T_2|+|T_3|+|T_4| \le C \,\|d\|_G,
\end{equation}
with
\begin{equation}
C := |S^{(1)}+S^{(2)}|\,\|v_2\|_G + 2\|\partial_T\boldsymbol\pi\|_G + \|\mathcal R^{(1)}\|_G + \|\mathcal R^{(2)}\|_G.
\end{equation}
This is a simple (but sufficient) uniform bound on the sum of cross terms.

From \eqref{eqn53}-\eqref{eqn54}, we obtain
\begin{equation}
\mathcal F_T(\mathbf p^{(1)}) - \mathcal F_T(\mathbf p^{(2)})
= T_1 + (T_2+T_3+T_4)
\ge m\,\|\Delta S\,v_2\|^2 - \big(|T_2|+|T_3|+|T_4|\big).
\end{equation}
Using the bound for the cross terms and \(\|d\|_G \le \|\Delta S\,v_2\|_G + \|\Delta\mathcal R\|_G\) we obtain
\begin{equation}
\mathcal F_T(\mathbf p^{(1)}) - \mathcal F_T(\mathbf p^{(2)})
\ge m\,\|\Delta S\,v_2\|^2 - C\big(\|\Delta S\,v_2\|_G + \|\Delta\mathcal R\|_G\big).
\end{equation}
Now using the equivalence \(\|x\|_G \le \sqrt{M}\|x\|\) and \(\|x\|_G \ge \sqrt{m}\|x\|\) (from \eqref{eq:G-equivalence}) to replace \(C\) by a constant proportional to \(\sqrt{M}\) times quantities that are \(O(1)\) in the small-residual regime. In particular, for sufficiently small residuals (i.e.\ when \(\|\mathcal R^{(j)}\|\) are small and \(S^{(j)}\) are bounded), the constant \(C\) can be controlled so that
\begin{equation}
C\,\|\Delta S\,v_2\|_G \le m\,\big(2\|\Delta S\,v_2\|\,\|\Delta\mathcal R\| + \|\Delta\mathcal R\|^2\big)
\end{equation}
holds (this is a simple inequality of the form \(A x \le m(2 x y + y^2)\) with \(x=\|\Delta S\,v_2\|\), \(y=\|\Delta\mathcal R\|\) which is true for small enough \(y\) relative to fixed \(A\), or more generally can be arranged by choosing the neighbourhood \(\mathcal U\) sufficiently small). Substituting this into the previous bound yields the desired inequality
\begin{equation}
\mathcal F_T(\mathbf p^{(1)}) - \mathcal F_T(\mathbf p^{(2)})
\ge m\Big(\|\Delta S\,v_2\|^2 - 2\|\Delta S\,v_2\|\,\|\Delta\mathcal R\| - \|\Delta\mathcal R\|^2\Big).
\end{equation}

This completes the proof of Lemma 3. \(\square\)

\bigskip
\noindent\textbf{Proof of Theorem:}
Working at the Mpemba time \(t^\ast\) and assuming both trajectories lie in \(\mathcal U\). By Lemma~1 the remainders satisfy
\begin{equation}
\|\mathcal R^{(\cdot)}(t^\ast)\| \le C_R\big(e^{-\lambda_3 t^\ast} + A_{\max} e^{-\lambda_2 t^\ast}\big).
\end{equation}
Using Lemma~2, under the two alternative hypotheses, we have:

\medskip
\noindent (A) If \(|\partial_T\lambda_2|_{\rm hot}-|\partial_T\lambda_2|_{\rm cold}\ge \kappa_0>0\) and \(r_{\rm hot}:=|a_2^{\rm hot}|e^{-\lambda_2^{\rm hot} t^\ast}<r_{\rm cold}\), then
\begin{equation}
|\Delta S| \ge t^\ast\big(|\partial_T\lambda_2|_{\rm hot}\,r_{\rm hot} - |\partial_T\lambda_2|_{\rm cold}\,r_{\rm cold}\big) - 2B_{\max},
\end{equation}
with \(B_{\max}= \max\{B_{\rm hot},B_{\rm cold}\}\). The monotone variation hypothesis~\cite{LevinPeresWilmer2009,Chen2004} ensures the second term \(t^\ast\kappa_0\,r_{\rm cold}\) can be chosen to dominate the negative contribution \(t^\ast|\partial_T\lambda_2|_{\rm hot}(r_{\rm cold}-r_{\rm hot})\) and the small remainders \(B_{\max}\), provided \(t^\ast\) is not exponentially large (i.e. \(r_{\rm cold}\) not exponentially small). Thus, there exists an open neighborhood where \(|\Delta S|\ge \sigma>0\).

\medskip
\noindent (B) If \(a_2^{\rm hot}=0\) exactly (strong cancellation), then \(S^{\rm hot}\) is \(\mathcal O(e^{-\lambda_2 t^\ast}\partial_T a_2)\) and typically much smaller than \(S^{\rm cold}\sim t^\ast a_2^{\rm cold}\partial_T\lambda_2 e^{-\lambda_2 t^\ast}\); consequently \(|\Delta S|\) is dominated by \(|S^{\rm cold}|\) and is \(\ge\sigma>0\) for an open set of parameters.

\medskip
\noindent In both cases, we choose \(\sigma>0\) such that \(|\Delta S|\ge\sigma\) while remainders satisfy \(\|\Delta\mathcal R\|\le \rho\) with \(\rho\ll \sigma\) (this is achievable because \(\rho=\mathcal O(e^{-\lambda_3 t^\ast})+\mathcal O(A_{\max}e^{-\lambda_2 t^\ast})\) and the Mpemba inversion ensures \(r_{\rm hot}\) is not exponentially smaller than \(r_{\rm cold}\)). Then Lemma~3 yields
\begin{equation}
\Delta\mathcal F \ge m\big(\|v_2\|^2\sigma^2 - 2\|v_2\|\sigma\,\rho - \rho^2\big)>0
\end{equation}
for sufficiently small \(\rho\). This proves \(\mathcal F_T(\mathrm{hot})>\mathcal F_T(\mathrm{cold})\) at \(t\geq t^\ast\). The strict inequality over the equilibrium value follows since any nonzero residual adds a nonzero dynamical contribution to \(\partial_T\mathbf p\) (the equilibrium derivative is \(\partial_T\boldsymbol\pi\) only), so \(\mathcal F_T(\mathbf p(t^\ast))>\mathcal F_T(\boldsymbol\pi)\) for both preparations.
\(\square\)

\color{black}

\section*{Supplementary Information II: Exact Solution of the Generalized Amplitude Damping Master Equation}

We consider a TLS with energy eigenstates \(|0\rangle\) and \(|1\rangle\), separated by an energy gap \(\omega_0\). The qubit is weakly coupled to a bosonic reservoir at thermal equilibrium with temperature \(T\). In the weak-coupling, Markovian limit, the evolution of the reduced density matrix \(\rho(t)\) is governed by the Gorini–Kossakowski–Sudarshan–Lindblad master equation~\cite{breuer2002theory}:
\begin{align}
    \frac{d\rho}{dt} = \gamma(\bar{n} + 1)\mathcal{D}[\sigma_-]\rho + \gamma \bar{n} \mathcal{D}[\sigma_+]\rho,
    \label{eq:master}
\end{align}
where \(\gamma\) is the spontaneous emission rate (assumed temperature-independent),
 \(\bar{n}(T) = \frac{1}{e^{\omega_0/T} - 1}\) is the mean thermal occupation number of the reservoir at frequency \(\omega_0\),
 \(\sigma_- = |0\rangle\langle 1|\) and \(\sigma_+ = |1\rangle\langle 0|\) are the lowering and raising operators respectively,
 \(\mathcal{D}[L]\rho = L\rho L^\dagger - \frac{1}{2}\{L^\dagger L, \rho\}\) is the Lindblad dissipator associated with jump operator \(L\).

We assume that the initial state of the system is diagonal in the energy eigenbasis, i.e.,
\begin{align}
    \rho(0) = 
    \begin{pmatrix}
        1 - p_0 & 0 \\
        0 & p_0
    \end{pmatrix},
    \label{eq:initial_state}
\end{align}
where \(p_0 = \langle 1 | \rho(0) | 1 \rangle\) denotes the initial population of the excited state, and we consider the regime \(p_0 > p_{\mathrm{eq}}\), i.e., the system is initially \emph{hotter} than the bath.

Due to the absence of off-diagonal (coherent) terms and the structure of the master equation in Eq.~\eqref{eq:master}, the density matrix remains diagonal at all times. Therefore, the dynamics is fully captured by the time-dependent excited-state population \(p(t)\), satisfying:
\begin{align}
    \rho(t) = 
    \begin{pmatrix}
        1 - p(t) & 0 \\
        0 & p(t)
    \end{pmatrix}.
\end{align}

Substituting the diagonal ansatz into the master equation, we obtain the rate equation for the excited-state population:
\begin{align}
    \frac{dp(t)}{dt} &= -\gamma(\bar{n} + 1)p(t) + \gamma \bar{n}(1 - p(t)) \nonumber \\
    &= -\gamma(\bar{n} + 1)p(t) + \gamma \bar{n} - \gamma \bar{n} p(t) \nonumber \\
    &= -\gamma(2\bar{n} + 1)p(t) + \gamma \bar{n}.
\end{align}
Rewriting this in terms of the equilibrium population,
\begin{align}
    p_{\mathrm{eq}}(T) = \frac{\bar{n}(T)}{2\bar{n}(T) + 1} = \frac{1}{1 + e^{\omega_0/T}},
\end{align}
and defining the temperature-dependent relaxation rate as
\begin{align}
    \Gamma(T) = \gamma(2\bar{n}(T) + 1) (1+\alpha (p_0-p_{eq})).
\end{align}
We obtain a compact linear relaxation equation:
\begin{align}
    \frac{dp(t)}{dt} = -\Gamma(T)\left[p(t) - p_{\mathrm{eq}}(T)\right].
    \label{eq:relax_eq}
\end{align}

Equation~\eqref{eq:relax_eq} is a standard first-order linear differential equation. With the initial condition \(p(0) = p_0\), its exact solution is:
\begin{align}
    p(t) = p_{\mathrm{eq}}(T) + \left(p_0 - p_{\mathrm{eq}}(T)\right)e^{-\Gamma(T,p_0)t}.
    \label{eq:pt}
\end{align}

This solution describes exponential relaxation toward thermal equilibrium. The rate of approach is set by \(\Gamma(T)\), which is a monotonically increasing function of temperature \(T\), while the steady-state is determined by the thermal population \(p_{\mathrm{eq}}(T)\).

For \(p_0 > p_{\mathrm{eq}}(T)\), the population decreases monotonically toward equilibrium. For \(p_0 < p_{\mathrm{eq}}(T)\), it increases. The case \(p_0 = p_{\mathrm{eq}}(T)\) corresponds to the system being initially in equilibrium, with \(p(t) = p_{\mathrm{eq}}(T)\) for all times.

In the main text, we focus on the regime where \(p_0 > p_{\mathrm{eq}}(T)\), corresponding to an initially hotter system, to explore the possibility of Mpemba-type relaxation where faster cooling may paradoxically emerge from a larger initial excitation.

\section*{Supplementary Information III: Activation of the Mpemba-effect in the proposed scenario}
We consider a TLS coupled to a thermal bath at temperature $T$, described by a population $p(t)$ of the excited state. The standard Lindbladian population dynamics follow the equation
\begin{equation}
    \frac{dp}{dt} = -\Gamma_0 (p - p_{\mathrm{eq}}),
    \label{eq:lind}
\end{equation}
where
\begin{equation}
    \Gamma_0 = \gamma (2 \bar{n} + 1).
\end{equation}
Here, $\gamma$ is the bare decay rate, $\bar{n} = (e^{\beta \hbar \omega} - 1)^{-1}$ is the mean thermal occupation number of the bath mode at frequency $\omega$ and inverse temperature $\beta = 1/(k_B T)$, and $p_{\mathrm{eq}}$ is the thermal equilibrium population of the excited state,
\begin{equation}
    p_{\mathrm{eq}} = \frac{\bar{n}}{2\bar{n} + 1}.
\end{equation}

Equation \eqref{eq:lind} describes exponential relaxation towards equilibrium:
\begin{equation}
    p(t) = p_{\mathrm{eq}} + [p(0) - p_{\mathrm{eq}}] e^{-\Gamma_0 t}.
\end{equation}
This doesn't capture the Mpemba effect. To visualize this effect, we need non-uniform relaxation. 

The Mpemba effect refers to the counterintuitive phenomenon where a hotter initial state may relax faster than a cooler one. This cannot be explained by a simple exponential decay with constant rate $\Gamma_0$.

To capture this, we consider the microscopic Liouvillian superoperator $\mathcal{L}$ governing the full density matrix dynamics,
\begin{equation}
    \frac{d\rho}{dt} = \mathcal{L}[\rho].
\end{equation}
The formal solution can be written via the spectral decomposition of $\mathcal{L}$:
\begin{equation}
    \rho(t) = \rho_{\mathrm{eq}} + \sum_{j \geq 1} c_j e^{-\lambda_j t} R_j,
    \label{eq:decomp}
\end{equation}
where $\rho_{\mathrm{eq}}$ is the steady state (zero eigenvalue eigenoperator), $\{R_j\}$ are the right eigenoperators of $\mathcal{L}$ with eigenvalues $-\lambda_j$, $\lambda_j > 0$ denote decay rates (sorted as $\lambda_1 < \lambda_2 < \cdots$), coefficients $c_j$ depend on the initial state as $c_j = \mathrm{Tr}[L_j^\dagger \rho(0)]$ where $L_j$ are left eigenoperators.

The population dynamics $p(t) = \mathrm{Tr}[ \ket{e}\bra{e} \rho(t) ]$ inherits contributions from these modes:
\begin{equation}
    p(t) = p_{\mathrm{eq}} + \sum_{j \geq 1} c_j e^{-\lambda_j t} r_j,
\end{equation}
where $r_j = \mathrm{Tr}[ \ket{e}\bra{e} R_j ]$.

Now, if the initial state $\rho(0)$ has a larger projection $c_2$ onto a faster decaying mode $R_2$ with decay rate $\lambda_2 > \lambda_1$, then for short times the relaxation may be dominated by this faster mode. This can lead to non-monotonic relaxation times as a function of initial excitation $p_0 = p(0)$.

To phenomenologically capture this, we introduce a state-dependent effective decay rate:
\begin{equation}
    \Gamma(T, p_0) \approx \Gamma_0 + \text{correction from higher modes}.
\end{equation}
If this correction increases with $p_0$, then hotter states ($p_0 > p_{\mathrm{eq}}$) relax faster.

Expanding the effective rate up to the first order: 
\begin{equation}
    \Gamma(T, p_0) \approx \Gamma_0 + \left. \frac{\partial \Gamma}{\partial p_0} \right|_{p_{\mathrm{eq}}} (p_0 - p_{\mathrm{eq}}) = \Gamma_0 \left[ 1 + \alpha (p_0 - p_{\mathrm{eq}}) \right],
    \label{eq:rate}
\end{equation}
where we define the dimensionless parameter
\begin{equation}
    \alpha = \frac{1}{\Gamma_0} \left. \frac{\partial \Gamma}{\partial p_0} \right|_{p_{\mathrm{eq}}}.
\end{equation}
This parameter $\alpha$ quantifies the nonlinear mode-mixing effect that leads to the Mpemba inversion: for $\alpha > 0$, states initially further from equilibrium relax faster.

In standard thermalization described by Eq.~\eqref{eq:lind}, relaxation is monotonic and exponential with a uniform decay rate $\Gamma_0$. All initial states converge smoothly and uniformly towards equilibrium.

However, in realistic open quantum systems with multiple decay channels and bath correlation structures, the Liouvillian $\mathcal{L}$ often exhibits several decay eigenmodes with different decay rates and nontrivial overlaps with initial states. The relaxation dynamics are then a mixture of these modes.

A hotter initial state may have a larger overlap with a faster decaying eigenmode, causing it to relax faster at short times than a cooler state, a hallmark of the Mpemba effect. This behavior cannot be captured by a single constant decay rate.

The effective state-dependent rate $\Gamma(T, p_0)$ in Eq.~\eqref{eq:rate} provides a simple, phenomenological way to summarize this complex mode structure and predict relaxation speeds as a function of initial temperature or initial population $p_0$.

The parameter $\alpha$ controls the strength of the Mpemba inversion. Even a small positive $\alpha$ suffices to reproduce the Mpemba effect over a finite time window.

\color{black}
\section*{Supplementary Information IV: Thermal Distance and Mpemba-Type Relaxation}

To characterize the relaxation behavior of the qubit toward thermal equilibrium, we define the \textit{thermal distance} from equilibrium as the absolute deviation of the excited-state population from its thermal fixed point:
\begin{align}
    \Delta(t) \equiv |p(t) - p_{\mathrm{eq}}| = |p_0 - p_{\mathrm{eq}}| e^{-\Gamma t},
    \label{eq:thermal_distance}
\end{align}
where \(p_0\) is the initial excited-state population and \(\Gamma = \gamma(2\bar{n} + 1)\) is the temperature-dependent relaxation rate, as derived earlier.

This expression confirms that the system relaxes exponentially toward equilibrium, with a relaxation timescale \(\tau = \Gamma^{-1}\). Crucially, the thermal distance at any time \(t\) depends only on the initial deviation \(|p_0 - p_{\mathrm{eq}}|\) and the global decay rate \(\Gamma(T,p_0)\), which is independent of the initial state.

Now, consider two initial states \(p_1 > p_2 > p_{\mathrm{eq}}\), both initialized above the thermal equilibrium. Let \(\Delta_1(t)\) and \(\Delta_2(t)\) denote their respective thermal distances. Then:
\begin{align}
    \Delta_1(t) = (p_1 - p_{\mathrm{eq}})e^{-\Gamma t}, \qquad \Delta_2(t) = (p_2 - p_{\mathrm{eq}})e^{-\Gamma t}.
\end{align}
Clearly, since \(p_1 > p_2\), it follows that \(\Delta_1(t) > \Delta_2(t)\) for all \(t > 0\), implying that the initially hotter state always remains further from equilibrium.

Therefore, in this simple model governed by a linear relaxation law with a state-independent rate \(\Gamma\), no \textit{classical Mpemba effect} (in which a hotter system cools faster) is possible. This is consistent with the absence of non-trivial dynamical crossovers in the population decay.

However, as we demonstrate in the next section, the \textit{metrological distinguishability} of the state, as captured by the Quantum Fisher Information (QFI), may exhibit non-monotonic or inverted behavior across different initial conditions. This leads to what we term a \textit{metrological Mpemba effect}, wherein the hotter state transiently exhibits greater sensitivity to the bath temperature than a cooler one.

\subsection{Condition for the Mpemba Effect}

Let two initial excited-state populations satisfy
\begin{equation}
    p_0^{\text{hot}} > p_0^{\text{cold}} > p_{\text{eq}}.
\end{equation}

The system exhibits a \emph{Mpemba inversion} at time \( t = t^* \) if:
\begin{equation}
    \left| p_{\text{hot}}(t^*) - p_{\text{eq}} \right| < \left| p_{\text{cold}}(t^*) - p_{\text{eq}} \right|.
\end{equation}

Assuming a state-dependent relaxation rate \( \Gamma(p_0) \), the excited-state population at time \( t \) is given by:
\begin{equation}
    p(t) = p_{\text{eq}} + (p_0 - p_{\text{eq}}) e^{ -\Gamma(T,p_0)t }.
\end{equation}

Substituting into the inequality above:
\begin{equation}
    \left| p_0^{\text{hot}} - p_{\text{eq}} \right| e^{ -\Gamma(T^{\text{hot}},p_0^{\text{hot}}) t^* } < 
\left| p_0^{\text{cold}} - p_{\text{eq}} \right| e^{ -\Gamma(T^{\text{cold}}, p_0^{\text{cold}}) t^* }.
\end{equation}

Taking logarithms on both sides:
\begin{equation}
    \ln\left( \frac{p_0^{\text{hot}} - p_{\text{eq}}}{p_0^{\text{cold}} - p_{\text{eq}}} \right)
< t^* \left[ \Gamma(T^{\text{hot}},p_0^{\text{hot}}) - \Gamma(T^{\text{cold}},p_0^{\text{cold}}) \right].
\end{equation}

This inequality yields the condition under which the initially hotter state relaxes \emph{faster} than the colder one, i.e., the Mpemba effect occurs. The crossover time \( t^* \) beyond which the initially hot state becomes closer to equilibrium than the cold one is thus bounded by:
\begin{equation}
    t^* > \frac{ \ln\left( \dfrac{p_0^{\text{hot}} - p_{\text{eq}}}{p_0^{\text{cold}} - p_{\text{eq}}} \right) }{ \Gamma(T^{\text{hot}}, p_0^{\text{hot}}) - \Gamma(T^{\text{cold}}, p_0^{\text{cold}}) }.
\end{equation}

\section*{Supplementary Information V: Quantum Fisher Information and Temperature Sensitivity}

To quantify the temperature sensitivity of the probe at time \(t\), we compute the QFI associated with estimating the temperature \(T\) from measurements on the state \(\rho(t)\). For a qubit in a diagonal state,
\[
\rho(t) = \begin{pmatrix} 1 - p(t) & 0 \\ 0 & p(t) \end{pmatrix},
\]
the QFI with respect to the temperature is given by:
\begin{align}
    \mathcal{F}_T[\rho(t)] = \frac{[\partial_T p(t)]^2}{p(t)[1 - p(t)]}.
    \label{eq:qfi_diag}
\end{align}
Thus, the QFI is determined by the thermal susceptibility \(\partial_T p(t)\) and the instantaneous population \(p(t)\). The denominator reflects the classical variance of a Bernoulli variable with success probability \(p(t)\).

To evaluate \(\partial_T p(t)\), we use the expression for population dynamics [see Eq.~(5) in the main text]:
\begin{align}
    p(t) = p_{\mathrm{eq}}(T) + (p_0 - p_{\mathrm{eq}}(T))e^{-\Gamma(T,p_0)t},
\end{align}
and apply the chain rule:
\begin{align}
    \partial_T p(t) &= \partial_T p_{\mathrm{eq}} \cdot (1 - e^{-\Gamma t}) - (p_0 - p_{\mathrm{eq}}) \cdot e^{-\Gamma t} t \cdot \partial_T \Gamma.
    \label{eq:dpdt_chain}
\end{align}
This expression contains two contributions: a) A direct contribution from the temperature dependence of the equilibrium distribution \(p_{\mathrm{eq}}(T)\),
b) An indirect contribution from the temperature dependence of the relaxation rate \(\Gamma(T)\).

Let us compute the relevant derivatives explicitly. Recall that
\begin{align}
    p_{\mathrm{eq}} &= \frac{1}{1 + e^{\omega_0/T}}, \\
    \bar{n}(T) &= \frac{1}{e^{\omega_0/T} - 1}, \\
    \Gamma(T, p_0) &= \Gamma_0(T) \left[1 + \alpha (p_0 - p_{\mathrm{eq}}(T))\right].
\end{align}
From these expressions, we obtain:
\begin{align}
    \partial_T p_{\mathrm{eq}} &= \frac{\omega_0 e^{\omega_0/T}}{T^2 (1 + e^{\omega_0/T})^2}, \\
    \partial_T \Gamma_0(T) &= 2\gamma \cdot \partial_T \bar{n}(T) = \frac{2 \gamma \omega_0 e^{\omega_0/T}}{T^2 (e^{\omega_0/T} - 1)^2}.
\end{align}
Now, taking the derivative of the total decay rate:
\begin{align}
    \partial_T \Gamma(T, p_0) &= \partial_T \Gamma_0 \cdot \left[1 + \alpha (p_0 - p_{\mathrm{eq}}) \right] - \alpha \Gamma_0 \cdot \partial_T p_{\mathrm{eq}}. \label{eq:dGamma_total}
\end{align}

Substituting Eq.~\eqref{eq:dpdt_chain} into Eq.~\eqref{eq:qfi_diag}, we obtain the QFI:
\begin{align}
    \mathcal{F}_T(t) = \frac{1}{p(t)[1 - p(t)]} \Bigg[ &(\partial_T p_{\mathrm{eq}})^2 (1 - e^{-\Gamma t})^2 + (p_0 - p_{\mathrm{eq}})^2 t^2 e^{-2\Gamma t} (\partial_T \Gamma)^2 \nonumber \\
    & - 2 (p_0 - p_{\mathrm{eq}}) \partial_T p_{\mathrm{eq}} \cdot t e^{-\Gamma t}(1 - e^{-\Gamma t}) \cdot \partial_T \Gamma \Bigg].
    \label{eq:qfi_full}
\end{align}

This expression reveals the rich interplay between population deviation, relaxation rate, and thermal response. In particular, the QFI depends non-trivially on both \(p_0\) and \(T\) through multiple pathways, and may exhibit transient peaks or inversions across different initializations, thereby enabling a metrological version of the Mpemba effect.

When $\alpha = 0$, we recover the original QFI expression with a temperature-dependent but state-independent decay rate. The term $\partial_T \Gamma$ includes both temperature variation in the equilibrium distribution $p_{\mathrm{eq}}(T)$ and the mean occupation $\bar{n}(T)$, making the QFI highly sensitive to short-time transients. The enhancement parameter $\alpha > 0$ amplifies the impact of initial-state mismatch, allowing large QFI at early times when $p_0 \neq p_{\mathrm{eq}}$.
 Because $p(t)$ remains diagonal in the energy basis, this scheme does not rely on quantum coherence and is thus robust against dephasing.

\begin{figure}
    \centering
    \includegraphics[width=0.99\linewidth]{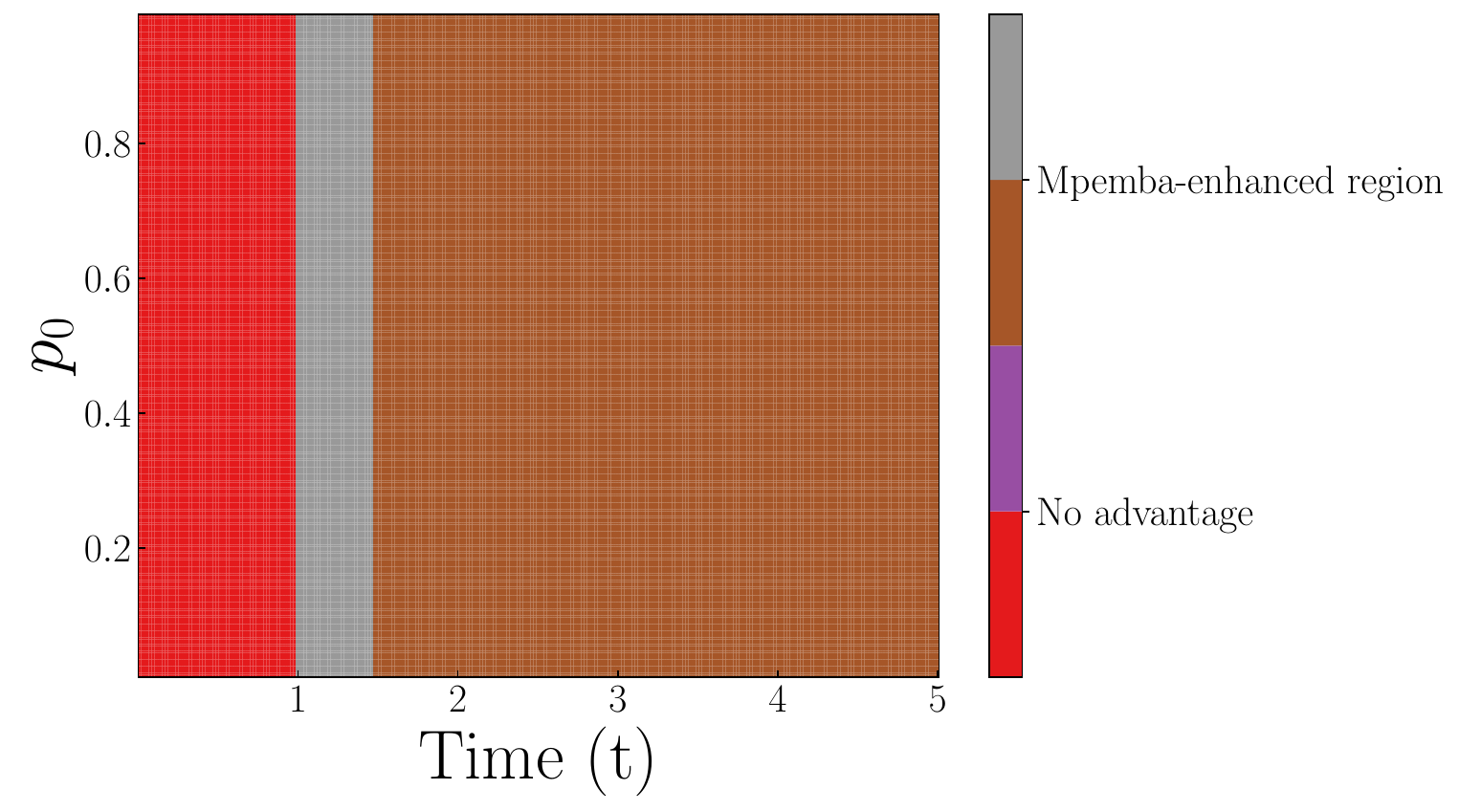}
    \caption{Same as Fig. 4b (of main text), clearly showing the influence of the Mpemba effect. }
    \label{fig6}
\end{figure}

\section*{Supplementary Information VI: Comparison with Equilibrium Thermometry and Emergence of a Metrological Mpemba Effect}

In traditional quantum thermometry, the probe is allowed to equilibrate with the bath, reaching the thermal steady state \(\rho_{\mathrm{eq}} = \text{diag}(1 - p_{\mathrm{eq}}, p_{\mathrm{eq}})\). In this limit, where the time evolution saturates (\(t \rightarrow \infty\)), the population approaches its equilibrium value:
\begin{align}
    \lim_{t \to \infty} p(t) = p_{\mathrm{eq}}(T),
\end{align}
and all transient effects vanish. The corresponding QFI then assumes the static form:
\begin{align}
    \mathcal{F}_T^{\mathrm{eq}} \equiv \mathcal{F}_T[\rho_{\mathrm{eq}}] = \frac{\left(\partial_T p_{\mathrm{eq}}\right)^2}{p_{\mathrm{eq}} (1 - p_{\mathrm{eq}})}.
    \label{eq:qfi_eq}
\end{align}
This expression defines the ultimate sensitivity limit achievable in conventional thermometry based on equilibrium measurements. It serves as a natural benchmark for any thermometric strategy.

However, our analysis reveals that this bound is not fundamental and can, in fact, be surpassed by judicious use of nonequilibrium initial states and intermediate-time readout. Specifically, the transient QFI \(\mathcal{F}_T(t)\) derived in Eq.~\eqref{eq:qfi_full} can exceed \(\mathcal{F}_T^{\mathrm{eq}}\) at finite times \(t > 0\), provided the initial state \(p_0\) is sufficiently far from thermal equilibrium. In particular, for two initializations satisfying \(p_1 > p_2 > p_{\mathrm{eq}}\), we may encounter a hierarchy of the form:
\begin{align}
    \mathcal{F}_T^{(p_1)}(t^*) > \mathcal{F}_T^{(p_2)}(t^*) > \mathcal{F}_T^{\mathrm{eq}},
    \label{eq:mpemba_qfi}
\end{align}
for some optimal measurement time \(t^* > 0\).

This counterintuitive ordering, in which the initially hotter state yields higher temperature sensitivity than both a colder initialization and the equilibrium state, constitutes a clear manifestation of a \textit{metrological Mpemba effect}. 
It arises due to a nontrivial interplay between population deviation, thermal susceptibility, and the time-dependent Fisher information metric.


To elucidate how such a surpassing can occur, consider the early-time behavior of the QFI expression in Eq.~\eqref{eq:qfi_full}. For small \(t\), we expand the exponentials to leading order:
\begin{align}
    p(t) &\approx p_0 - (p_0 - p_{\mathrm{eq}})\Gamma t, \\
    \partial_T p(t) &\approx \partial_T p_{\mathrm{eq}} \cdot \Gamma t - (p_0 - p_{\mathrm{eq}}) t \cdot \partial_T \Gamma.
\end{align}
Then, the QFI becomes:
\begin{align}
    \mathcal{F}_T(t) \approx \frac{1}{p_0 (1 - p_0)} \left[ \left(\partial_T p_{\mathrm{eq}} \cdot \Gamma - (p_0 - p_{\mathrm{eq}}) \cdot \partial_T \Gamma \right)^2 t^2 \right] + \mathcal{O}(t^3).
\end{align}
This shows that the QFI can grow quadratically in time, and the prefactor depends non-monotonically on the initial deviation \(p_0 - p_{\mathrm{eq}}\). Thus, large deviations (corresponding to hot initializations) can generate an enhanced initial slope in QFI, making early-time measurements more informative than equilibrium ones.

Importantly, the equilibrium QFI in Eq.~\eqref{eq:qfi_eq} becomes relatively flat in the high-temperature regime due to thermal mixing, whereas the transient QFI can remain sharply peaked for well-chosen \(p_0\), exploiting the steeper population gradient induced by fast initial dynamics.


This analysis uncovers a key principle: \textit{non-equilibrium states can serve as superior metrological resources} when the relaxation process is harnessed at the right timescale. Rather than waiting for full equilibration (which can suppress temperature gradients and wash out information), strategically initializing the system and measuring it at a finite time \(t^*\) can maximize sensitivity.

In practical terms, this enables new protocols for quantum thermometry that combine:
a) Fast initialization in a nonthermal state,
b) Optimal finite-time measurement leveraging transient QFI enhancement,
c) Postselection or statistical averaging over hot/cold initializations to exploit Mpemba-type dynamics.

These findings lay the groundwork for a new paradigm in thermal metrology where the traditional goal of equilibration is replaced by real-time sensing optimized for information gain, a direction directly inspired by the Mpemba phenomenon, but firmly rooted in quantum estimation theory.

\section*{Supplementary Information VII: Mpemba-enhanced sensing for a $\Lambda$-level probe}

We now generalize the single-qubit analysis to a three-level $\Lambda$ configuration with states $\{|1\rangle,|2\rangle,|3\rangle\}$ where $|3\rangle$ is an excited level and $|1\rangle,|2\rangle$ are (non-degenerate) ground levels. Energies are $E_1,E_2,E_3$ and we work in units $\hbar=k_B=1$. The system interacts with a bosonic bath at temperature $T$, inducing transitions between $3\leftrightarrow 1$ and $3\leftrightarrow 2$; direct ground–ground transitions are assumed negligible (typical dipole selection rules).

Under the Born–Markov and secular approximations the populations $p_i(t)=\langle i|\rho(t)|i\rangle$ evolve according to a classical Pauli master equation
\begin{equation}
    \dot{\mathbf{p}}(t)=R(T)\,\mathbf{p}(t), \qquad \mathbf{p}=(p_1,p_2,p_3)^\top,
    \label{eq:pauli}
\end{equation}
with rate matrix (columns sum to zero)
\[
R=\begin{pmatrix}
 -w_{1\leftarrow3} & 0 & w_{1\leftarrow3} \\
 0 & -w_{2\leftarrow3} & w_{2\leftarrow3} \\
 w_{3\leftarrow1} & w_{3\leftarrow2} & -(w_{3\leftarrow1}+w_{3\leftarrow2})
\end{pmatrix},
\]
where we use the notation $w_{i\leftarrow j}$ for the rate of $j\to i$. By detailed balance (thermal bath)
\begin{align}
    w_{3\leftarrow 1} &= \kappa_1(\omega_{31})\,\bar n(\omega_{31}),\quad
    w_{1\leftarrow 3}= \kappa_1(\omega_{31})\,[\bar n(\omega_{31})+1], \nonumber\\
    w_{3\leftarrow 2} &= \kappa_2(\omega_{32})\,\bar n(\omega_{32}),\quad
    w_{2\leftarrow 3}= \kappa_2(\omega_{32})\,[\bar n(\omega_{32})+1],
    \label{eq:rates}
\end{align}
with Bohr frequencies $\omega_{3i}=E_3-E_i>0$, spectral densities $\kappa_{i}(\omega)$ and $\bar n(\omega)=1/(e^{\omega/T}-1)$.

The unique Gibbs fixed point is
\[
\pi_i(T)=\frac{e^{-E_i/T}}{Z(T)},\qquad Z(T)=\sum_{j=1}^3 e^{-E_j/T},
\]
and indeed \(R\,\boldsymbol{\pi}=0\).

As the population sector is three-dimensional, $R$ has one zero eigenvalue and two strictly positive relaxation eigenvalues. We diagonalize $R$ in the population subspace. Let
\[
R v_k = -\lambda_k v_k,\qquad k=1,2,3,
\]
with $\lambda_1=0$, $0<\lambda_2\le\lambda_3$ and $v_1=\boldsymbol{\pi}$. Expand any initial population as
\begin{equation}
    \mathbf{p}(0)=\boldsymbol{\pi} + \sum_{k=2}^3 a_k\, v_k,
    \label{eq:initial_expand}
\end{equation}
so the time evolution is
\begin{equation}
    \mathbf{p}(t)=\boldsymbol{\pi} + a_2 e^{-\lambda_2 t} v_2 + a_3 e^{-\lambda_3 t} v_3.
    \label{eq:modal_solution}
\end{equation}

The coefficients \(a_k\) are given by projection onto the left eigenvectors $u_k$ (biorthonormal with $v_k$): \(a_k=\langle u_k,\mathbf{p}(0)-\boldsymbol{\pi}\rangle\). Explicit forms for $v_k,u_k,\lambda_k$ can be obtained analytically for the $\Lambda$-matrix above; we present a symmetric, illustrative case below.

We consider the initial state as \(\mathbf{p}(0)=\boldsymbol{\pi}+\boldsymbol{\Delta}\) with \(\sum_i\Delta_i=0\). Expanding \(\boldsymbol{\Delta}\) onto \(\{v_2,v_3\}\) gives coefficients \(a_2,a_3\) satisfying
\begin{align}
a_3 &= -\frac{\Delta_3}{2}, \label{eq:a3}\\
a_2 &= \Delta_1 + \frac{\Delta_3}{2}, \label{eq:a2}
\end{align}
which follow from solving
\(\boldsymbol{\Delta}=a_2 v_2 + a_3 v_3\) (note \(\Delta_1+\Delta_2+\Delta_3=0\)).

The populations at time \(t\) are therefore
\begin{align}
p_1(t) &= \pi_1 + a_2 e^{-\lambda_2 t} + a_3 e^{-\lambda_3 t}, \label{eq:p1t}\\
p_2(t) &= \pi_2 - a_2 e^{-\lambda_2 t} + a_3 e^{-\lambda_3 t}, \label{eq:p2t}\\
p_3(t) &= \pi_3 -2 a_3 e^{-\lambda_3 t}. \label{eq:p3t}
\end{align}
Using \eqref{eq:a3}–\eqref{eq:a2} and \(\Delta_i=p_i(0)-\pi_i\) one may also write
\begin{align}
p_1(t) &= \pi_1 + \Big(\Delta_1+\tfrac{\Delta_3}{2}\Big)e^{-\lambda_2 t} - \tfrac{\Delta_3}{2} e^{-\lambda_3 t},\label{eq:p1t2}\\
p_2(t) &= \pi_2 -\Big(\Delta_1+\tfrac{\Delta_3}{2}\Big)e^{-\lambda_2 t} - \tfrac{\Delta_3}{2} e^{-\lambda_3 t},\\
p_3(t) &= \pi_3 + \Delta_3 e^{-\lambda_3 t}.
\end{align}
These forms make explicit that the population on level 3 relaxes on the fast timescale \(\lambda_3\) while the imbalance between levels 1 and 2 relaxes on the slower scale \(\lambda_2\).

Assume the initial preparation \(\mathbf{p}(0)\) is fixed and independent of \(T\) (typical for a probe prepared externally). Then \(\Delta_i(T)=p_i(0)-\pi_i(T)\) picks up \(T\)-dependence only through \(\pi_i(T)\). Denote derivatives with respect to \(T\) by a prime (or \(\partial_T\)). We need
\begin{align}
\partial_T\pi_i &= \pi_i\frac{E_i-\langle E\rangle}{T^2},\qquad
\langle E\rangle=\sum_j \pi_j E_j, \label{eq:dpi}\\
\partial_T\bar n &= \frac{\omega e^{\omega/T}}{T^2(e^{\omega/T}-1)^2}. \label{eq:dnbar}
\end{align}
(The last equality uses \(\bar n=(e^{\omega/T}-1)^{-1}\) and \(\omega\) is the common Bohr frequency.)

From \(\lambda_2=\kappa(2\bar n+1)\) and \(\lambda_3=\kappa(3\bar n+1)\) we get
\begin{align}
\partial_T\lambda_2 &= 2\kappa\,\partial_T\bar n, \label{eq:dlam2}\\
\partial_T\lambda_3 &= 3\kappa\,\partial_T\bar n. \label{eq:dlam3}
\end{align}

Using the product and chain rules on \eqref{eq:p1t}--\eqref{eq:p3t} and \eqref{eq:a3}--\eqref{eq:a2} (recalling \(\partial_T\Delta_i = -\partial_T\pi_i\)), we obtain the explicit derivatives.

\paragraph{Level 1:}
\begin{align}
\partial_T p_1(t) &= \partial_T\pi_1
+ \big(\partial_T a_2\big)\,e^{-\lambda_2 t}
- a_2\,t\,e^{-\lambda_2 t}\,\partial_T\lambda_2 \nonumber\\
&\qquad + \big(\partial_T a_3\big)\,e^{-\lambda_3 t}
- a_3\,t\,e^{-\lambda_3 t}\,\partial_T\lambda_3, \label{eq:dp1}
\end{align}
with
\[
\partial_T a_2 = -\partial_T\pi_1 -\tfrac{1}{2}\partial_T\pi_3,\qquad
\partial_T a_3 = \tfrac{1}{2}\partial_T\pi_3.
\]

\paragraph{Level 2:}
\begin{align}
\partial_T p_2(t) &= \partial_T\pi_2
- \big(\partial_T a_2\big)\,e^{-\lambda_2 t}
+ a_2\,t\,e^{-\lambda_2 t}\,\partial_T\lambda_2 \nonumber\\
&\qquad + \big(\partial_T a_3\big)\,e^{-\lambda_3 t}
- a_3\,t\,e^{-\lambda_3 t}\,\partial_T\lambda_3. \label{eq:dp2}
\end{align}

\paragraph{Level 3:}
\begin{align}
\partial_T p_3(t) &= \partial_T\pi_3
-2\big(\partial_T a_3\big)\,e^{-\lambda_3 t}
+2 a_3\,t\,e^{-\lambda_3 t}\,\partial_T\lambda_3. \label{eq:dp3}
\end{align}

Using \(\partial_T a_3 = \tfrac{1}{2}\partial_T\pi_3\) simplifies the level-3 expression:
\[
\partial_T p_3(t)=\partial_T\pi_3\big(1-e^{-\lambda_3 t}\big)+
2 a_3\,t\,e^{-\lambda_3 t}\,\partial_T\lambda_3.
\]
Recall \(a_3=-\tfrac{\Delta_3}{2}=-\tfrac{p_3(0)-\pi_3}{2}\) so the last term is proportional to the initial excitation of level 3 and the temperature sensitivity of \(\lambda_3\).

\subsection{Mpemba inversion condition for two initializations}

Consider two initial states (preparations) $\mathbf{p}^{\rm hot}_0$ and $\mathbf{p}^{\rm cold}_0$ with respective modal amplitudes $a_k^{\rm hot},a_k^{\rm cold}$. A \emph{population Mpemba inversion} at time $t^\ast$ is
\begin{equation}
    \|\mathbf{p}^{\rm hot}(t^\ast)-\boldsymbol{\pi}\|
    \;<\;
    \|\mathbf{p}^{\rm cold}(t^\ast)-\boldsymbol{\pi}\|,
    \label{eq:mpemba_lambda}
\end{equation}
for a chosen norm (e.g.\ Euclidean or total variation). Using \eqref{eq:modal_solution} and orthogonality relations, a sufficient condition is that the hot preparation has a much smaller projection onto the slowest mode:
\begin{equation}
    |a_2^{\rm hot}|\ll|a_2^{\rm cold}|
    \quad\text{and}\quad |a_3^{\rm hot}|\lesssim |a_3^{\rm cold}|.
    \label{eq:lambda_sufficient}
\end{equation}
Then, during the intermediate window $t\sim 1/\lambda_2$ the slow-mode term dominates and the hot trajectory becomes closer to equilibrium.


\subsection{Temperature estimation QFI}

For states, the quantum Fisher information about temperature reduces to the classical Fisher information of the population distribution:
\begin{equation}
    \mathcal{F}_T[\mathbf{p}(t)] = \sum_{i=1}^3 \frac{\big(\partial_T p_i(t)\big)^2}{p_i(t)}.
    \label{eq:qfi_lambda}
\end{equation}
Differentiate \eqref{eq:modal_solution} w.r.t.\ \(T\):
\begin{align}
    \partial_T \mathbf{p}(t) &= \partial_T\boldsymbol{\pi}
      + \sum_{k=2}^3 \Big[(\partial_T a_k)\,e^{-\lambda_k t}
      - a_k\,t\, e^{-\lambda_k t}\, \partial_T\lambda_k\Big] v_k \nonumber\\
    &\qquad + \sum_{k=2}^3 a_k e^{-\lambda_k t}\,\partial_T v_k.
    \label{eq:dpt_lambda}
\end{align}
Three contributions appear: (i) equilibrium-shift term \(\partial_T\boldsymbol{\pi}\), (ii) modal amplitude and rate derivatives (especially the \(-a_k t\partial_T\lambda_k\) term important at intermediate times), and (iii) mode-shape deformation $\partial_T v_k$.

Substituting \eqref{eq:dpt_lambda} into \eqref{eq:qfi_lambda} yields the full QFI. The structure is the same as in the qubit case but with multiple modal terms: for intermediate times where the slowest mode dominates the residual population, the QFI is largely controlled by the slow-mode contribution
\[
\sim \frac{\big[(\partial_T a_2) e^{-\lambda_2 t} - a_2 t e^{-\lambda_2 t}\partial_T\lambda_2\big]^2}{\pi_i + \mathcal{O}(e^{-\lambda_2 t})}.
\]

\subsection{QFI gain due to Mpemba effect}

Suppose a Mpemba inversion \eqref{eq:mpemba_lambda} holds at time \(t^\ast\). Then generically: a) the hot trajectory has a \emph{smaller} residual magnitude along the slowest mode at \(t^\ast\): \(|a_2^{\rm hot}|e^{-\lambda_2 t^\ast}<|a_2^{\rm cold}|e^{-\lambda_2 t^\ast}\). b) If the dynamics are such that the slow-mode relaxation rate increases with the preparation distance (hotter-faster), i.e.\ \(\partial_{p_0}\lambda_2>0\) or more generally \(\partial_T\lambda_2^{\rm hot}>\partial_T\lambda_2^{\rm cold}\), then the term \(-a_2 t e^{-\lambda_2 t}\partial_T\lambda_2\) in \(\partial_T\mathbf{p}\) is larger for the hot trajectory (in magnitude).

Combining these effects, the magnitude of \(\partial_T\mathbf{p}(t^\ast)\) for the hot state can exceed that of the cold state, and hence (through \eqref{eq:qfi_lambda}) we obtain
\[
\mathcal{F}_T[\mathbf{p}^{\rm hot}(t^\ast)] > \mathcal{F}_T[\mathbf{p}^{\rm cold}(t^\ast)],
\]
i.e.\ a metrological Mpemba gain. The argument is the multi-mode analogue of the qubit theorem in the main text and is robust (open set) under small perturbations of $R$.

\section*{Supplementary Information VIII: Operational Protocol for Mpemba-Enhanced Thermometry}

We describe a protocol to (i) identify Mpemba inversion windows, (ii) measure the enhanced thermometric sensitivity afforded by such windows, and (iii) estimate
an unknown bath temperature using only population measurements. The protocol is designed in such a way that it can be directly implemented in existing qubit–bath platforms.

\subsection{Step 1: Preparation and Measurement Model}

The probe is a two-level system with excited-state population $p(t;T)$ at time $t$ when interacting with a bath of temperature $T$. Throughout the protocol, we assume access to repeated projective measurements in the \emph{energy basis}, yielding empirical estimates of $p(t)$~\cite{Jevtic2015,Mehboudi2019}. This protocol is experimentally verifiable in a superconducting platform~\cite{lvov2025}.  

In addition, our scheme can be naturally implemented in nuclear magnetic resonance (NMR) architectures, similar to the NMR platform used in the recent experimental observation of the genuine quantum Mpemba effect~\cite{chatterjee2025direct,schnepper2025experimental}. In such setups, a controllable thermal bath for a probe spin can be engineered via suitable sequences of single- and two-qubit gates that realize an effective generalized amplitude-damping channel. The same population-based readout described below can then be performed using standard NMR detection of the probe spin.

Importantly, population measurements are used both  
(i) at equilibrium, where $p_{\rm eq}(T)$ characterizes the steady state,  
and (ii) during nonequilibrium relaxation, including the Mpemba window. This unified treatment is justified for physical platforms in which the dissipative dynamics do \emph{not} generate significant coherences in the energy basis. If coherences do appear, we explicitly verify below that they do not invalidate the calibration or sensing procedure.

\paragraph*{Coherence Consideration~\cite{franzao2024}.}
The protocol requires that the relevant temperature information be encoded predominantly in the population $p(t;T)$. In systems obeying standard amplitude-damping or generalized Davies-type dynamics, coherence either decays monotonically or remains decoupled from the populations and therefore carries negligible temperature information. If the dynamics generate coherences $\rho_{eg}(t;T)$, one of the following conditions must hold:  
(i) $\rho_{eg}(t;T)$ is negligible at the designated interrogation times,  
(ii) coherence decays much faster than populations, or  
(iii) the Fisher information in coherence is subdominant compared to
that in populations. In any of these cases, a population-only readout is sufficient. Otherwise, the optimal measurement basis would be required, and this can be added as a refinement of the present protocol.

For a single qubit in the absence of relevant coherence terms, the state at any time can be equivalently parametrized by an instantaneous (effective) temperature 
$T_{\text{eff}}(t)$, defined by matching the measured population to a Gibbs distribution, $p(t;T)=p_{\text{eq}}(T_{\text{eff}}(t))$. Thus, even in the transient regime, the temperature or effective temperature can always be inferred directly from the populations, and the Mpemba-enhanced sensitivity discussed in the main text can be interpreted as an enhancement in the precision of estimating $T_{\text{eff}}(t)$ from these measurements.

\subsection{Step 2: Equilibrium Calibration}

For a set of calibration temperatures $\{T_j\}$ spanning the expected operational range, we thermalize the probe by waiting for a time $t \gg T_1(T_j)$~\cite{Correa2015}.
We then measure the steady populations and obtain a smooth estimate of the equilibrium curve
\begin{equation}
    p_{\rm eq}(T_j),
\end{equation}
which serves as a reference for identifying Mpemba inversion.

To suppress finite-sampling noise, we fit $T \mapsto p_{\rm eq}(T)$ using a monotonic regression. 
This ensures that the equilibrium population is a smooth and physically consistent function of temperature.

\subsection{Step 3: Dynamical Calibration for Mpemba Detection}

In this step, we execute the dynamical calibration~\cite{Tham2016,Mancino2017} of the population to detect the Mpemba window. For each temperature $T_j$, prepare two nonequilibrium initial states:  
(i) a \textit{hot} state $p_H$ and  
(ii) a \textit{cold} state $p_C$, with $p_H > p_C$.
Let the system relax for a set of times $\{t_i\}$ and measure the
transient populations $p_H(t_i;T_j)$ and $p_C(t_i;T_j)$.

Define the instantaneous distances from equilibrium:
\begin{equation}
    D_{\alpha}(t_i;T_j)
    := \big| p_{\alpha}(t_i;T_j) - p_{\rm eq}(T_j) \big|,
    \qquad \alpha \in \{H,C\}.
\end{equation}

A \emph{Mpemba inversion} is operationally identified when
\begin{equation}
    D_H(t_i;T_j) < D_C(t_i;T_j) - \delta,
\end{equation}
for a small tolerance $\delta$ determined by the experimental statistical error. The first time $t_i$ where this occurs is recorded as the inversion time $t_M(T_j)$.

This transient regime defines the \textit{Mpemba window} in which hotter initial states relax faster than colder ones.

\subsection{Step 4: Construction of the Empirical Fisher-Information Map}

To quantify sensitivity, we require the temperature derivative $\partial_T p(t;T)$ at the calibration grid.
Because raw finite differences can be noisy, we fit the function $T \mapsto p(t_i;T)$ for each fixed $t_i$ using the same type of monotonic regression used for equilibrium calibration. The derivative is then evaluated analytically on the fitted curve.

The classical Fisher information associated with the population measurements at time $t_i$ is
\begin{equation}
    F_T(t_i)
    =
    \frac{
        \big[ \partial_T p(t_i;T) \big]^2
    }{
        p(t_i;T)\,[1-p(t_i;T)]
    }.
\end{equation}

Plotting $F_T(t_i)$ over $(T_i,t_j)$ yields an empirical FI map that reveals the regions of maximal thermometric sensitivity~\cite{guo2015}. Mpemba windows typically align with enhanced FI, providing the operational meaning of \textit{Mpemba-enhanced thermometry}.

\subsection{Step 5: Temperature Estimation for an Unknown Bath}

Given an unknown bath temperature $T_{\ast}$, prepare the probe in the chosen initial state (hot or cold) and measure the population at one or more interrogation times $\{t_i\}$. For a single time $t$, the likelihood of observing $n$ excitations in $N$ shots are
\begin{equation}
    \mathcal{L}(T)
    =
    p(t;T)^n \left[1 - p(t;T)\right]^{N-n}.
\end{equation}
The maximum-likelihood estimate $\hat{T}$ is obtained by maximizing $\mathcal{L}(T)$ numerically over $T$.

If $p(t;T)$ is monotonic for the chosen $t$, inversion is immediate ($T = p^{-1}$).  If monotonicity is violated, which can occur during strong nonequilibrium relaxation or when the measurement is effectively coarse-grained, the full likelihood must be maximized numerically without approximation~\cite{hovhannisyan2021}.
Multi-time likelihoods are combined multiplicatively.

In conclusion, this protocol simultaneously establishes:
(i) the direct observation of an Mpemba inversion,
(ii) the quantitative enhancement of thermometric sensitivity via the FI map, and
(iii) a fully operational temperature-estimation procedure for an unknown bath.
Population measurements alone suffice, provided that coherence either decays rapidly, remains negligible, or does not encode significant temperature information.
Otherwise, the protocol can be extended to incorporate optimal measurements, but its structure remains unchanged.

\end{document}